\documentclass[a4paper,11pt]{article}

\usepackage{amsmath,amssymb}
\usepackage[dvipdfmx]{graphicx}
\usepackage{bm}
\usepackage{diagbox}
\usepackage{mathptmx}
\usepackage{subcaption}
\usepackage{braket}
\usepackage{hyperref}
\usepackage{color}

\title{
\vspace{-20mm}
{\normalsize \hfill KANAZAWA-25-04}\\
\vspace{20mm}
Phase structure analysis of 2d lattice CP(1) model with $\theta$ term using tensor renormalization group method}

\author{
Hayato Aizawa\thanks{\texttt{h\_aizawa@hep.s.kanazawa-u.ac.jp}} , Shinji Takeda, and Yusuke Yoshimura\\
{\it Institute for Theoretical Physics, Kanazawa University, Kanazawa 920-1192, Japan}
}

\begin{document}

\maketitle

\begin{abstract}
We investigate the phase structure of a two-dimensional lattice CP(1) model with a $\theta$ term.
In particular, we aim to identify a critical region expected to exist along a $\theta=\pi$ line.
To explore the phase structure non-perturbatively and avoid the sign problem, we employ the tensor renormalization group method.
We make two improvements compared to previous tensor network studies.
The first improvement involves refining the initial tensor.
Specifically, we construct it using a quadrature method, which achieves higher accuracy compared to the conventional approach.
The second improvement consists of analyzing the phase structure using the information of the conformal field theory, namely the central charge and the scaling dimensions, which can be accessed relatively easily via the tensor renormalization group method.
Thanks to these improvements, we identify both the onset of the critical region, $\beta_{\rm c}=0.5952(8)$ and its universality class as the SU(2)${}_{k=1}$ Wess-Zumino-Witten model.
\end{abstract}
\newpage
\section{Introduction}
Quantum Chromodynamics (QCD) allows the presence of a topological $\theta$ term in its Lagrangian \cite{Callan:1976je}.
The presence of the $\theta$ term, however, renders the associated Boltzmann weight a complex number, 
which leads to the sign problem in MC simulations and makes the investigation of this term particularly difficult \cite{PhysRevLett.94.170201}.
In nature, the value of the $\theta$ parameter is known to be extremely small, which is referred to as the strong CP problem \cite{CREWTHER1979123,PhysRevLett.124.081803}.
Despite this, investigating QCD with a finite $\theta$ term remains of theoretical interest, 
as it can help us better understand nonperturbative and topological aspects of the strong interaction.

Tensor renormalization group (TRG) methods have attracted attention as a promising numerical tool for addressing the sign problem and have been applied to various models exhibiting this problem \cite{PhysRevD.90.074503,10.1093/ptep/ptv022,Kawauchi:2017dnj,Kadoh:2018hqq, PhysRevD.105.054507,Akiyama:2020soe,10.1093/ptep/ptac014}.
In addition, several algorithmic improvements have been proposed in recent years \cite{PhysRevLett.99.120601, PhysRevB.86.045139, PhysRevLett.115.180405, PhysRevB.97.045111, PhysRevB.102.054432, PhysRevB.105.L060402, Nakayama:2023ytr}.
Nevertheless, the computational cost of TRG method grows rapidly with the spacetime dimension and the number of internal degrees of freedom, making the analysis of QCD in four-dimensional spacetime particularly challenging.

\begin{figure}[t!]
    \centering
    \includegraphics[width=0.45\columnwidth]{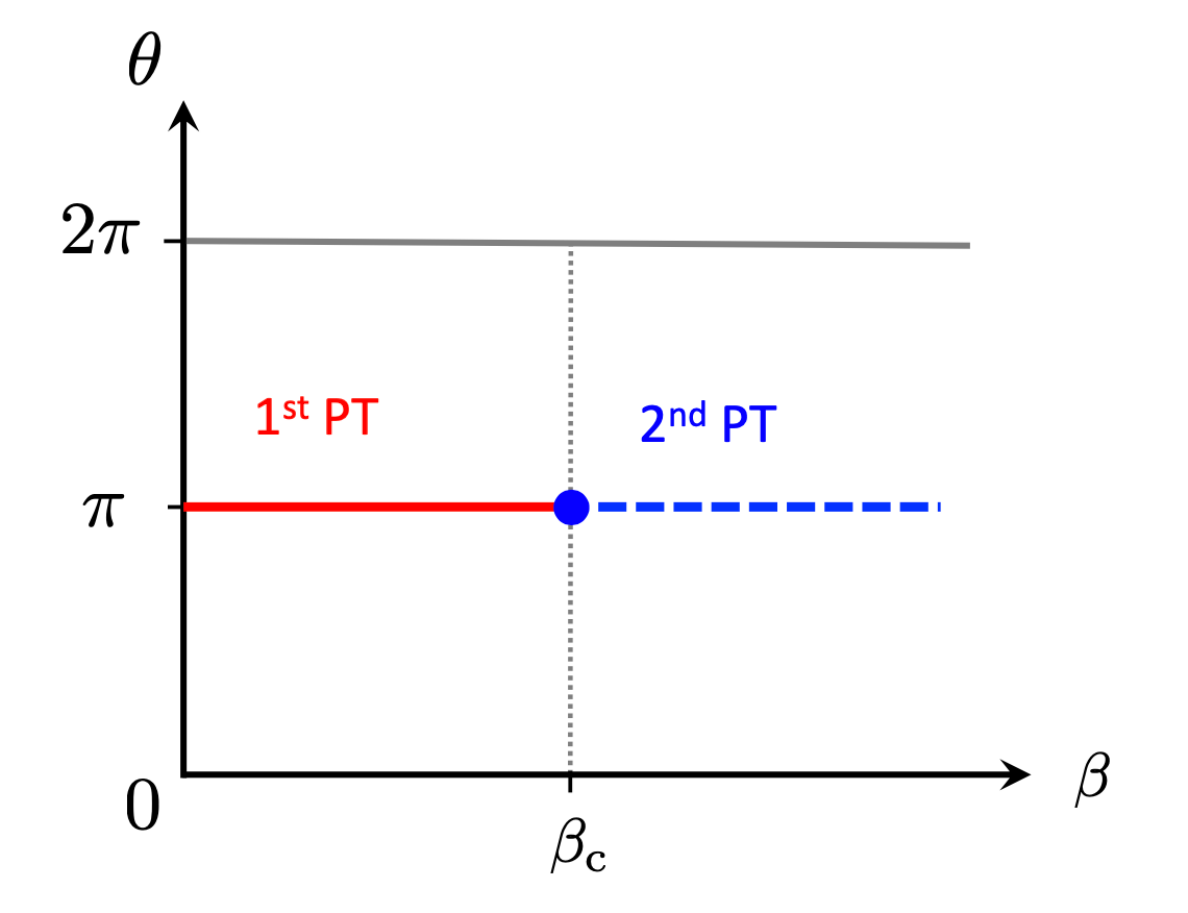}
   \caption{Expected phase structure of 2d lattice CP(1) model with $\theta$ term.
   At $\theta=\pi$ and in the strong coupling region ($\beta\sim0$), the phase transition is first-order.
   As $\beta$ increases, the transition weakens, and eventually the system enters the critical region, whose existence is anticipated from the Haldane conjecture.
   At the onset $\beta=\beta_{\rm c}$ of the critical region, the universality is expected to be the SU(2)$_{k=1}$ WZW model.
   }
   \label{fig:phase_structure_CP1_pre}
\end{figure}

It is therefore natural to study a toy model of QCD, namely CP($N-1$) model, which shares essential features with QCD such as asymptotic freedom, confinement, and the $\theta$ term. 
The CP($N-1$) model is characterized by two parameters, the inverse coupling constant $\beta$ and $\theta$, and its expected phase structure on the lattice is illustrated in Fig.~\ref{fig:phase_structure_CP1_pre}.
Strong coupling analyses \cite{PhysRevLett.53.637, PhysRevD.55.3966} have shown that, at $\theta=\pi$ in the strong coupling region, the phase transition is of first-order,
and that it becomes weaker as $\beta$ increases.
For sufficiently large $\beta$, however, the strong coupling analysis ceases to be valid, and numerical methods are required.

On the other hand, in the case of CP(1)($N=2$), there are qualitative predictions for the phase structure thanks to its relation to the Haldane conjecture.
The conjecture states that the ground state of a one-dimensional spin-$s$ antiferromagnet drastically changes depending on the value of $s$, which is shown by using the effective model O(3) nonlinear $\sigma$ model \cite{HALDANE1983464,PhysRevLett.50.1153}.
In this correspondence, the spin $s$ of the antiferromagnet maps to the $\theta$ parameter of the O(3) model, and it is found that the ground state is gapless at $\theta=\pi$, while it has a gap at $\theta=0$.
When the ground state is gapless, the system is at a critical point where its universality class is reflected in the critical exponents.
Analytical studies of the O(3) model using non-Abelian bosonization, etc., \cite{PhysRevB.36.5291,PhysRevLett.66.2429} have shown that its universality class is SU(2)$_{k=1}$ Wess-Zumino-Witten (WZW) model \cite{Wess:1971yu,Witten:1983ar}.
Numerical studies of the lattice O(3) model have further confirmed the existence of the corresponding critical point at $\theta=\pi$ \cite{PhysRevLett.75.4524,PhysRevD.77.056008,PhysRevD.86.096009}.
Furthermore, the study using D-theory regularization also produced results consistent with this prediction \cite{Caspar:2022llo}.
Since the CP(1) model and the O(3) nonlinear $\sigma$ model are equivalent in the continuum limit, it is expected that the lattice CP(1) model also exhibits the same criticality at $\theta=\pi$.
In fact, numerical calculation, which employs a sign-problem-free formulation of the CP(1) model based on the Villain-type regularization of the U(1) gauge theory coupled to 2-flavor Higgs fields,
has found the expected critical points \cite{Sulejmanpasic:2020lyq}.

Nevertheless, numerical studies of the lattice CP(1) model have not yet identified the expected critical point.
In a MC study using the analytic continuation method, a critical point was observed around $\beta=0.5$, but the corresponding critical exponent did not agree with that of the SU(2)$_{k=1}$ WZW model \cite{PhysRevLett.98.257203}.
This discrepancy may be attributed to the sign problem; to clarify this point, a method free from the sign problem, such as TRG method, is required.
In the first TRG study \cite{Kawauchi:2017dnj}, a critical region was reported to emerge from $\beta\approx0.4$, whereas a subsequent study with a different coarse-graining algorithm \cite{PhysRevD.105.054507} found no evidence of a critical point up to $\beta=1.1$.
In this way, contradictions remain regarding the existence of a critical region even when using TRG methods.

In this study, we reanalyze the phase structure of the lattice CP(1) model using TRG method, with two key improvements over the previous studies. 
The first improvement concerns the construction of the initial tensor.
In earlier works \cite{Kawauchi:2017dnj,PhysRevD.105.054507}, it was constructed by truncating the character expansions \cite{PhysRevD.55.3966,10.1143/ptp/93.1.161}. 
However, the character expansion, particularly for the $\theta$ term is known to converge poorly.
As a result, the initial tensor obtained in this way is subject to large truncation errors, which  were not carefully evaluated.
To avoid relying on this expansion, we instead employ the quadrature method \cite{Kadoh:2018hqq,GENZ2003187,Kuramashi:2019cgs} to construct a new initial tensor and, moreover, we explicitly examine its error.
The second improvement is the analysis of the phase structure using information of the conformal field theory (CFT), namely a central charge and scaling dimensions.
In previous studies \cite{Kawauchi:2017dnj,PhysRevD.105.054507}, the finite size scaling of the topological susceptibility was used to diagnose the nature of phase transition.
However, it is known that usual finite size scaling is strongly affected by logarithmic corrections near the critical point of the SU(2)$_{k=1}$ WZW model \cite{PhysRevB.48.16814},
which is the same universality class of the BKT transition for the two-dimensional XY model \cite{PhysRevLett.55.1355}.
A solution to this problem was proposed in \cite{KNomura_1994}, which relies on the level spectroscopy using CFT data,
and has recently been further developed by combining TRG method \cite{PhysRevB.104.165132}.
The new analysis accurately determine the BKT transition point in the two-dimensional XY model as a crossing point of the scaling dimensions.
In the current paper, we also adopt this method to analyze the phase structure of the lattice CP(1) model.
As a result of these improvements, we are able to identify both the location of the onset of the critical region and its universality as the SU(2)$_{k=1}$ WZW model.

The rest of the paper is organized as follows.
In section \ref{sec:CP1}, we fix the notations used throughout the paper.
The action and the partition function of CP(1) model on the square lattice are defined in this section.
In section \ref{sec:TN}, we present a tensor network representation of the partition function of the model, and examine the accuracy of the initial tensor.
Numerical results for the phase structure analysis, including CFT data, are shown in section \ref{sec:result}.
Our concluding remarks are given in section \ref{sec:conclusion}.
Appendix \ref{ch} provides the details of constructing the initial tensor using the character expansion.
In appendix \ref{genz}, we describe the quadrature method on hyperspheres.
Appendix \ref{ex} presents the exact result of the free energy on a $2\times2$ lattice.
In appendix \ref{sec:symmetry}, we discuss to what extent the important symmetries are maintained in numerical calculations.

\section{CP(1) model on 2d lattice}
\label{sec:CP1}
The action of the CP(1) model with a $\theta$ term on a two-dimensional square lattice $x\in\{(x_1,x_2)|x_1,x_2=0,1,\dots,L-1\}$ is given by
\begin{equation}
    S=-2\beta \sum_{x,\mu}
    \left[
    z^\dag(x)z(x+\hat\mu)e^{iA_\mu(x)}
    +
    z^\dag(x+\hat\mu)z(x)e^{-iA_\mu(x)}
    \right]
    -
    \frac{\theta}{2\pi}
    \sum_x \log U_p(x),
\label{eqn:action}
\end{equation}
where the lattice spacing is set to $a=1$, and $\hat\mu$ denotes the unit vector in the $\mu$-direction ($\mu=0,1$). Periodic boundary conditions (PBC) are imposed in both directions.
Here, $z(x)$ is a two-component complex scalar field,
\begin{equation}
    z(x)=
    \left(
    \begin{array}{c}
    z_1(x)
    \\
    z_2(x)
    \end{array}
    \right)
    \in\mathbb{C}^2,
\end{equation}
subject to the constraint
\begin{equation}
    |z(x)|^2=1.
\end{equation}
The plaquette loop $U_p(x)$ is expressed in terms of the U(1) gauge field $A_\mu(x)$ as 
\begin{equation}
    U_p(x)=\exp\left\{i\left[
A_0(x)+A_1(x+\hat0)-A_0(x+\hat1)-A_1(x)
    \right]\right\}.
\end{equation}

The partition function is defined as
\begin{equation}
    Z(\beta,\theta)
    =\left(\prod_x\int dz(x)\right)
    \left(\prod_{x,\mu}\int_{-\pi}^{\pi}\frac{dA_\mu(x)}{2\pi}\right)
    e^{-S},
    \label{eq:Z}
\end{equation}
where the integral measure of the complex scalar field $dz(x)$ is given by
\begin{equation}
    \int dz(x)=C_z\int_{\mathbb{C}^2} d\mathrm{Re}(z_1(x))d\mathrm{Im}(z_1(x))d\mathrm{Re}(z_2(x))d\mathrm{Im}(z_2(x))\delta(|z(x)|^2-1),
\label{eqn:original_measure}
\end{equation}
and the normalization constant $C_z$ is defined such that
$   \int dz(x)=1$.

Using the constraint and gauge transformation, the complex scalar field $z$ can be written more concisely. First, taking the constraint on $z$ into account,
it can be parametrized by three angular variables $\phi_1\in(0,\pi)$ and $\phi_2,\phi_3\in(0,2\pi)$,
\begin{equation}
    z(x)
    =e^{i\phi_3(x)}\left(
        \begin{array}{c}
        \cos(\frac{\phi_1(x)}{2})
        \\
        \sin(\frac{\phi_1(x)}{2})e^{i\phi_2(x)}
        \end{array}
    \right).
\end{equation}
Furthermore, by performing the gauge transformation for $A_\mu$
\begin{equation}
    A_\mu(x)^\prime=A_\mu(x)+\phi_3(x+\hat{\mu})-\phi_3(x),
\end{equation}
the phase $\phi_3$ can be absorbed into the gauge field.
As a result, the scalar field can be expressed solely in terms of the two variables $(\phi_1,\phi_2)$:
\begin{equation}
    z(x)
    =\left(
        \begin{array}{c}
        \cos(\frac{\phi_1(x)}{2})
        \\
        \sin(\frac{\phi_1(x)}{2})e^{i\phi_2(x)}
        \end{array}
    \right).
\end{equation} 
Hence, $\phi_3$ can be trivially integrated out, and the integral measure reduces to that on the two-dimensional unit sphere
\begin{equation}
    \int dz =\frac{1}{4\pi}\int_0^\pi d\phi_1 \sin\phi_1 \int_0^{2\pi} d\phi_2.
\label{eqn:measure_S2}
\end{equation}
Note that the original measure in Eq.~\eqref{eqn:original_measure} is defined for the three-dimensional unit sphere.
In the following, however, we employ the same notation for the two-dimensional unit sphere.

For later convenience, we define $H$ and $Q$ as 
\begin{equation}
    H_{zz^\prime A}:=\exp\left(2\beta
     \left[
    z^\dag z^\prime e^{iA}
    +
    z^{\prime\dag}ze^{-iA}
    \right]
    \right),
    \label{eq:H}
\end{equation}
\begin{eqnarray}
    Q_{AA^\prime A^{\prime\prime}A^{\prime\prime\prime}}&:=&\exp\left(\frac{\theta}{2\pi}\log \left[
    e^{i(A+A^\prime-A^{\prime\prime}-A^{\prime\prime\prime})}
    \right]\right).
    \label{eq:Q}
\end{eqnarray}
Then, the partition function can be rewritten in terms of $H$ and $Q$ as
\begin{equation}
    Z(\beta,\theta)
    =\left(\prod_x\int dz(x)\right)
    \left(\prod_{x,\mu}\int_{-\pi}^{\pi}\frac{dA_\mu(x)}{2\pi}\right)
    \prod_{x,\mu}H_{z(x),z(x+\hat\mu),A_\mu(x)}
    \prod_{x}Q_{A_0(x),A_1(x+\hat0),A_0(x+\hat1),A_1(x)}.
    \label{eq:ZHQ}
\end{equation}

\section{Tensor network representation}
\label{sec:TN}
To perform an analysis using TRG method, one first has to express the partition function $Z$ in terms of a tensor network.
The tensor appearing in this representation is referred to as the initial tensor.
In previous studies \cite{Kawauchi:2017dnj,PhysRevD.105.054507}, the initial tensor was constructed by truncating the character-like expansion for $H$ \cite{PhysRevD.55.3966} and 
the character expansion for $Q$ \cite{ 10.1143/ptp/93.1.161}, 
respectively (see Appendix \ref{ch} for details of these expansions).
However, it is known that the expansion for $Q$ converges poorly.
Hence, the initial tensor obtained from these expansions is likely to suffer from large truncation errors, which were not carefully estimated.
Here, we propose a new initial tensor by discretizing the field integration using the quadrature method \cite{Kadoh:2018hqq,GENZ2003187,Kuramashi:2019cgs}.

First, we approximate the integral of the gauge field using the Gauss-Legendre quadrature method,
\begin{equation}
    \int_{-1}^{1}dy\,f(y)
    \approx
    \sum_{a=1}^{N_{\rm A}}
	w_a f(y_a),    
\end{equation}
where $y_a$ is the $a$-th sample point in the Gauss-Legendre quadrature, $w_a$ is the corresponding weight, and $N_{\rm A}$ is the number of sample points.
By performing the variable transformation to the gauge field $A=\pi\,y$, the integral of a function $g(A)$ can be approximated as
\begin{equation}
    \int_{-\pi}^{\pi}    \frac{dA}{2\pi}\,g(A)
    \approx \sum_{a=1}^{N_\mathrm{A}} W_{a}^{(\rm{A})} g(A_a),
    \label{eq:Na}
\end{equation}
where $W_{a}^{(\rm{A})}$ and $A_a$ are given by
\begin{equation}
    W_{a}^{(\rm{A})}=\frac{1}{2}w_a,\quad A_a=\pi\,y_a.
\end{equation}

Next, as explained in Eq.~\eqref{eqn:measure_S2}, the integration for the scalar field is carried out over the two-dimensional unit sphere $U_2=\{\bm{y}|\bm{y}\in\mathbb{R}^3,\bm{y}^2=1\}$, for which a convenient quadrature formula is available \cite{GENZ2003187} (see Appendix \ref{genz} for details of the general hypersphere $U_{n-1}$ case),
\begin{equation}
    \int_{U_2}d\sigma f(\bm{y})\approx\sum_{i=1}^{N_{\rm{z}}} w_i f(\bm{y}_i),
\end{equation}
where $d\sigma$ is the normalized measure on $U_2$, $w_i$ are the weights given in Eq.~\eqref{eqn:Wz},
$\bm{y}_i$ are the sample points, and $f(\bm{y}_i)$ is given in Eq,~\eqref{eqn:sample_fx}.
Here, $N_{\rm z}$ denotes the number of sample points, which is determined by the quadrature order $m\in\mathbb{N}$ and a free parameter $\mu$ of the quadrature method (see Eq.~\eqref{eqn:Nz_general} for general $U_{n-1}$ case with $\mu\neq0$). 
The actual values $N_{\rm{z}}$ as a function of $m$ for $U_2$ and $\mu\neq0$ are summarized in Table~\ref{tab:spherical_m_N}.
In our calculation, we set $\mu=10^{-6}$.
\begin{table}[t!]
    \centering
    \caption{The number of sample points $N_{\rm z}$ as a function of the quadrature order $m$ for $U_2$ and $\mu\neq0$.}
    \begin{tabular}{c|cccccccc}
    \hline
    $m$ & 1 & 2 & 3 & 4 & 5 & 6 & 7 & 8 \\
    \hline
    $N_{\rm z}$ & 24 & 48 & 80 & 120 & 168 & 224 & 288 & 360 \\
    \hline
    \end{tabular}
    \label{tab:spherical_m_N}
\end{table}
To apply the quadrature, a coordinate $\bm{y}$ on $U_2$ should be mapped to the complex scalar field $z$,
\begin{equation}
    z(\bm{y})
    =\left(
        \begin{array}{c}
        \cos(\frac{\phi_1}{2})
        \\
        \sin(\frac{\phi_1}{2})e^{i\phi_2}
        \end{array}
    \right)
    =\left(
        \begin{array}{c}
        \sqrt{\frac{1+y_3}{2}}
        \\
        \sqrt{\frac{1-y_3}{2({y_1}^2+{y_2}^2)}}(y_1+iy_2)
        \end{array}
    \right),
    \label{zx}
\end{equation}
via the phase variables $\phi_1$ and $\phi_2$ with $(y_1,y_2,y_3)=(\sin\phi_1\cos\phi_2,\sin\phi_1\sin\phi_2,\cos\phi_1)$.
By combining the transformation with the quadrature, the integral of the complex scalar field is now approximated as
\begin{equation}
	\int dz \,\,g(z)\approx
    \sum_{i=1}^{N_{\rm{z}}} W_i^{(\rm{z})} g(z_i),
    \label{eq:Nz}
\end{equation}
with
\begin{equation}
    W_i^{(\rm{z})}=w_i,\quad z_i=z(\bm{y}_i).
\end{equation}

\begin{figure}
    \centering
    \includegraphics[width=0.9\columnwidth]{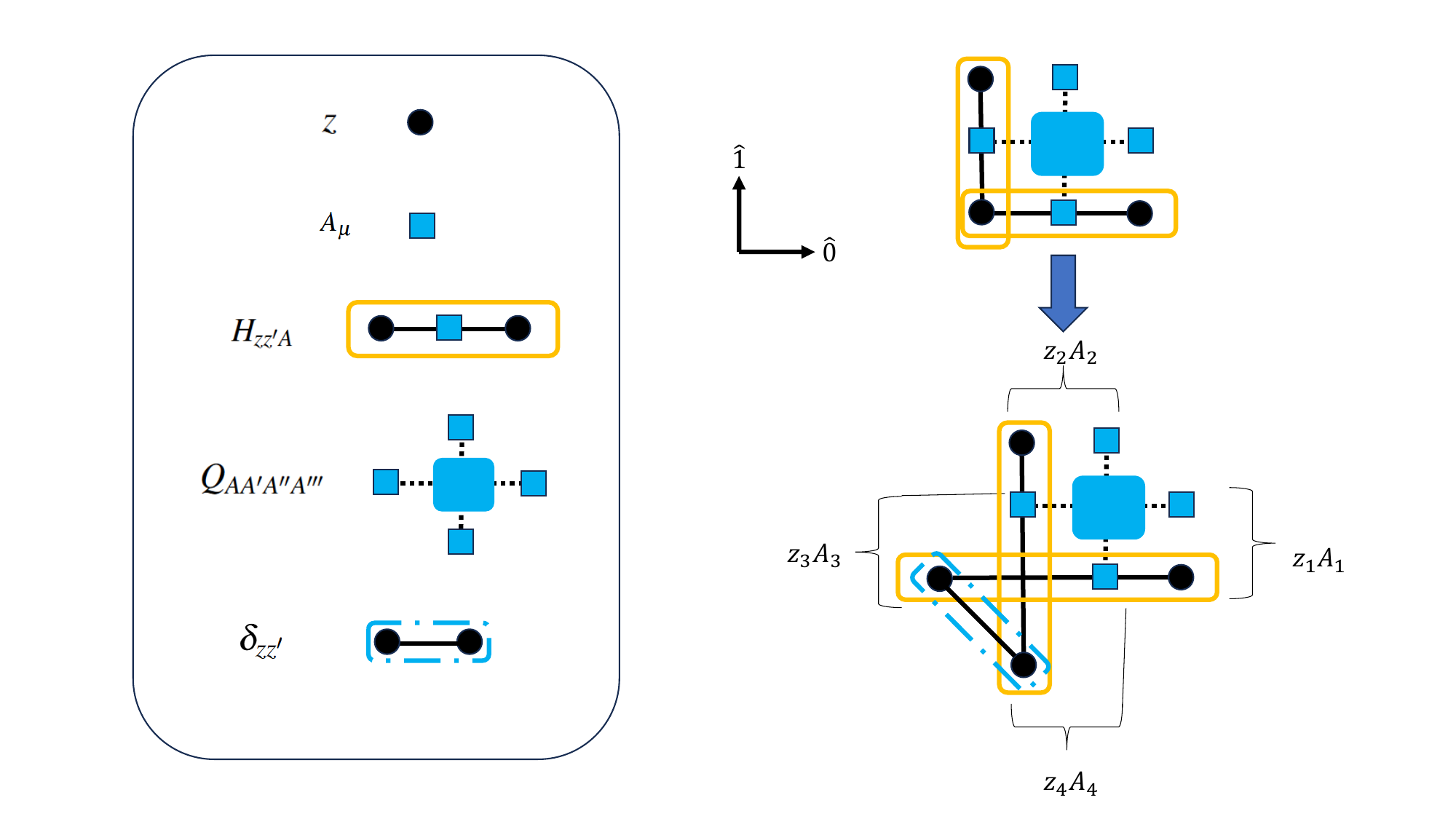}
    \caption{The initial tensor constructed using the quadrature method. The left box summarizes some building blocks: the complex scalar field $z$ (the gauge field $A_\mu$) is denoted by black circle (blue square).
    $H$ in Eq.~\eqref{eq:H}, $Q$ in Eq.~\eqref{eq:Q}, and a Kronecker delta are shown as the circles and the squares connected by the lines.
    In the right panel, two $H$s and $Q$ with an inserted Kronecker delta form the initial tensor, which has four paired indices $(z_iA_i)$ with $i=1,2,3,4$.}
    \label{fig:tensor_quad}
\end{figure}

As seen from Eq.~\eqref{eq:ZHQ}, the partition function is expressed as a tiling of the lattice, where each unit consists of $HHQ$ as shown in
the top-right panel of Fig.~\ref{fig:tensor_quad}.
By applying the quadrature formulas in Eq.~\eqref{eq:Na} and \eqref{eq:Nz} for $A$ and $z$, the partition function can be written as a collection of tensor contractions.
Note that in order to ensure the network involves only nearest-neighbor contraction, Kronecker deltas are inserted as shown in the bottom-right panel of Fig.~\ref{fig:tensor_quad}.
The resulting tensor network is then given as,
\begin{equation}
Z\approx\left(\prod_x\sum_{i_x}^{N_{\rm z}}\sum_{{i_x}^\prime}^{N_{\rm z}}\right)\left(\prod_{x,\mu}\sum_{a_{x,\mu}}^{N_{\rm A}}\right)
    \prod_x T_{(i_{x+\hat{0}},a_{x+\hat{0},1})(i^{\prime}_{x+\hat{1}},a_{x+\hat{1},0})(i_x,a_{x,1})({i_x}^\prime,a_{x,0})},
\end{equation} 
where the tensor $T$ is defined as
\begin{equation}
    T_{(z_1A_1)(z_2A_2)(z_3A_3)(z_4A_4)}
    =
    W_{z_4}^{({\rm z})}
    \sqrt{W_{A_1}^{({\rm A})}
    W_{A_2}^{({\rm A})}
    W_{A_3}^{({\rm A})}
    W_{A_4}^{({\rm A})}}
    \delta_{z_3,z_4}
    H_{z_3,z_1,A_4}
    H_{z_4,z_2,A_3}
    Q_{A_4,A_1,A_2,A_3}.
    \label{eqn:tensor_naive}
\end{equation}
Note that the integers $i_x,a_{x,\mu}$ appearing in the sums of Eq.~\eqref{eq:Na} and \eqref{eq:Nz} become the indices of the tensor.

\begin{figure}[tb]
    \centering
\begin{minipage}{0.49\columnwidth}
    \centering
    \includegraphics[width=0.9\columnwidth]{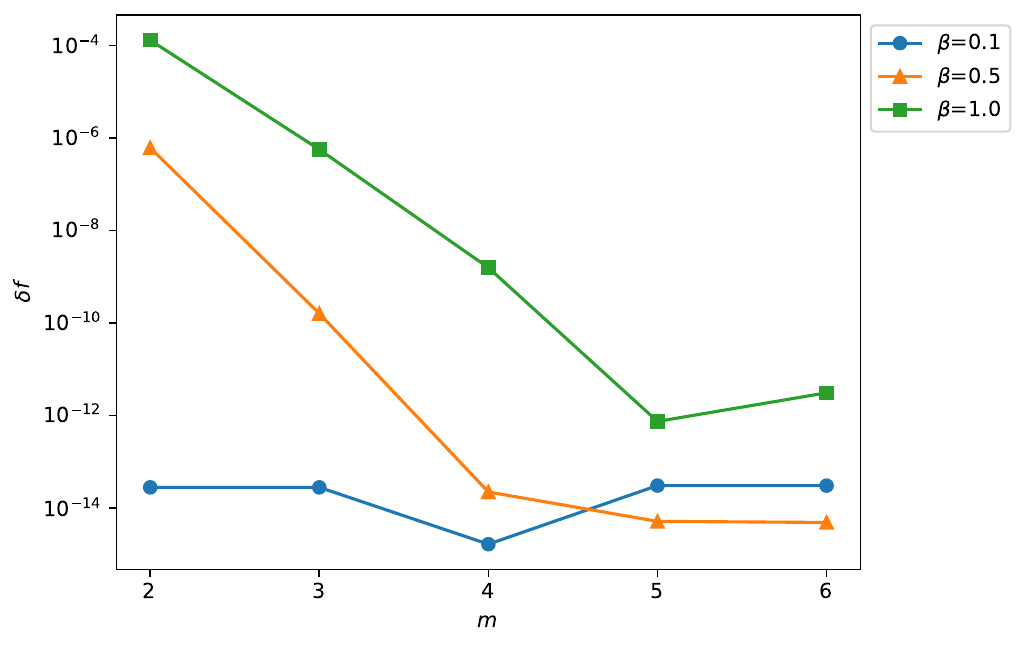}
    \subcaption{}
    \label{fig:cp12by2_m_df}
\end{minipage}
\begin{minipage}{0.49\columnwidth}
    \centering
    \includegraphics[width=0.9\columnwidth]{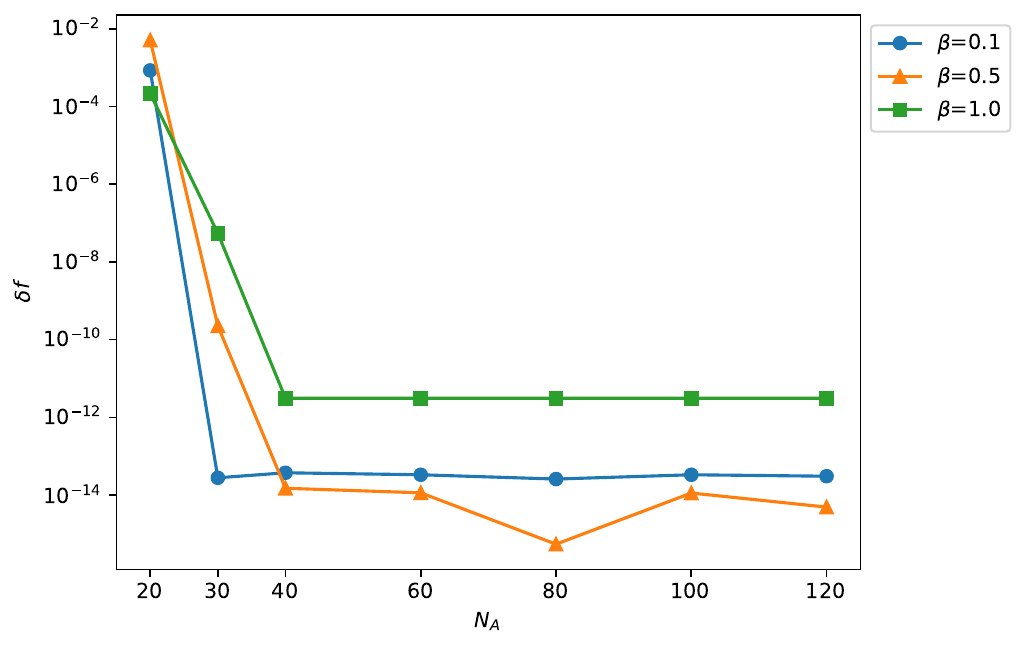}
    \subcaption{}
    \label{fig:cp12by2_Na_df}
\end{minipage}
   \caption{Relative error of the free energy on $2\times2$ lattice, defined in Eq.~\eqref{eqn:error1}, with $\theta=\pi$, as a function of the quadrature order.
   Panel (a) shows the dependence on $m$ for fixed $N_{\rm{A}}=120$, while (b) shows the dependence on $N_{\rm A}$ for $m=6$ ($N_{\rm z}=224$).
   The blue circles, the orange triangles, and the green squares represent data for $\beta=0.1$, $0.5$, and $1.0$, respectively. 
   }
   \label{fig:cp12by2_quad_df}
\end{figure}

Here, let us verify the error introduced by the quadrature methods.
To this end, we compute a free energy $f_{\rm tensor}$ on $2\times2$ lattice by contracting the four initial tensors in Eq.~\eqref{eqn:tensor_naive} without any coarse-graining\footnote{In practice, to compute $f_{\rm tensor}$ on $2\times2$ lattice we do not explicitly make the tensor in Eq.~\eqref{eqn:tensor_naive}. Instead, we discretize the integration over the gauge field in Eq. \eqref{integrateA} using the quadrature formula in Eq. \eqref{eq:Na}, and evaluate the remaining parts following the procedure given in Appendix \ref{ex}.}, and
compare it with the exact free energy $f_{\rm exact}$.
On such a small lattice, a brute-force exact calculation is feasible (see Appendix \ref{ex} for details and the numerical results).
The relative error of the free energy is defined as
\begin{equation}
    \delta f = \frac{|f_{\rm tensor} - f_{\rm exact}|}{|f_{\rm exact}|}.
\label{eqn:error1}
\end{equation}
Figure~\ref{fig:cp12by2_quad_df} shows the relative error as a function of the quadrature order, $m$ and $N_{\rm A}$.
We perform the calculation at $\theta=\pi$, which is expected to be the point with the largest error.
As a result, we find that in order to achieve $\delta f<10^{-4}$ for $\beta = 0.1$ -- $1.0$, the quadrature orders should satisfy $m\geq3$ and $N_{\rm A}\geq30$.
This indicates that the required bond dimension for the initial tensor 
is of order $O(N_{\rm z}\times N_{\rm A})\sim O(10^3$ -- $10^4)$,
which renders actual computations for larger volumes infeasible.

To avoid such a difficulty, we compress the tensor in Eq.~\eqref{eqn:tensor_naive} and reduce its dimensionality
by employing a HOTRG-like technique \cite{PhysRevB.86.045139}.
This procedure provides a projector $P^{(1)}$ and
reduces the dimension of the index, for example $(z_1A_1)$, of the tensor from $N_{\rm z} \times N_{\rm A}$ to $D_{\rm c} \sim O(10$ -- $100)$.
This is achieved by minimizing the following cost function\footnote{When making the projector, it is advantageous to use the initial tensor expressed in terms of the character expansion in Eq.~\eqref{eqn:tensor_ch}, since in that case the projector can be fully factorized into the complex scalar part and the gauge field. The computational cost of making such a block-diagonalized projector is relatively low and it turns out to be comparable to that of the coarse-graining procedure.},
\begin{equation}
    \left|T_{(z_1A_1)(z_2A_2)(z_3A_3)(z_4A_4)}-\sum_{z,A}\sum_{i=1}^{D_{\rm c}}T_{(zA)(z_2A_2)(z_3A_3)(z_4A_4)}P^{(1)}_{zA,i}P^{(1)\dag}_{z_1A_1,i}\right|^2.
\end{equation}
Exploiting the translational symmetry of the tensor networks, the index $(z_3 A_3)$ is compressed using $P^{(1)\dag}$.
Similarly, by making $P^{(2)}$ for the direction of $(z_2 A_2)$, we finally obtain the compressed initial tensor in which all bond dimensions are reduced to $D_{\rm c}$,\footnote{
Using reflection symmetry, $P^{(2)}$ can be obtained from $P^{(1)}$.
}
\begin{equation}
    T_{i_1i_2i_3i_4}^{({\rm compressed})}=\sum_{z_1,z_2,z_3,z_4,A_1,A_2,A_3,A_4}T_{(z_1A_1)(z_2A_2)(z_3A_3)(z_4A_4)}
    P^{(1)}_{z_1A_1,i_1}P^{(2)}_{z_2A_2,i_2}P^{(1)\dag}_{z_3A_3,i_3}P^{(2)\dag}_{z_4A_4,i_4}
    \label{eqn:tensor_init}
\end{equation}
with $i_k=1,2,\dots,D_{\rm c}$ ($k=1,2,3,4$).

\begin{figure}[t!]
    \centering
    \includegraphics[width=0.6\columnwidth]{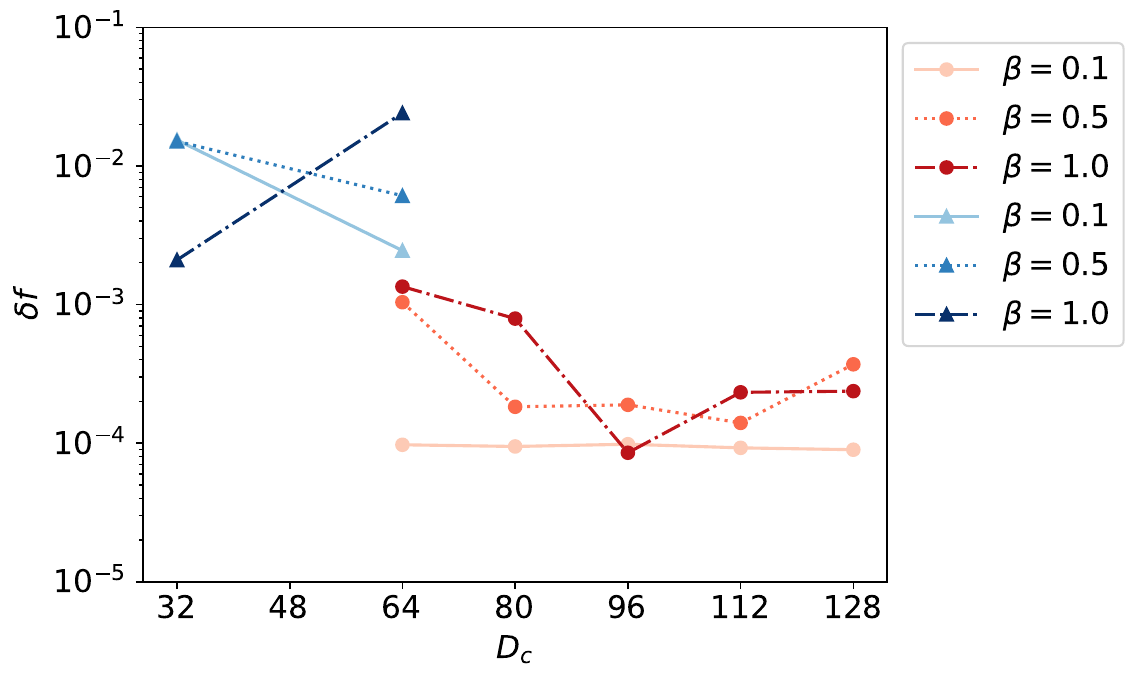}
   \caption{Comparison of the relative error for the quadrature method and the character expansion for $\theta=\pi$ and $\beta=0.1$ -- $1$.
		For the quadrature method, the compressed initial tensor in Eq.~\eqref{eqn:tensor_init} is used and its results are represented by the red circles as a function of $D_{\rm c}$.
   The quadrature orders are fixed to $m=6$ and $N_{\rm A}=120$.  
The blue triangles represent the results for the initial tensor in Eq.~\eqref{eqn:tensor_ch} using the conventional character expansion ($D_{\rm c}=\chi_{\rm z}\times\chi_{\rm A}=16\times\chi_{\rm A}$ where $\chi_{\rm z}$ and $\chi_{\rm A}$ are defined in Eq.~\eqref{eqn:chi_z} and \eqref{eqn:chi_A} in Appendix \ref{ch}).
   }
   \label{fig:cp12by2_Dc_df}
\end{figure}
To evaluate the error introduced by the compression, we again employ the relative error on $2\times2$ lattice for the compressed initial tensor in Eq.~\eqref{eqn:tensor_init}.
The results are shown in Fig.~\ref{fig:cp12by2_Dc_df}, where it can be seen that the error of the compressed tensor is of order $O(10^{-4})$.
On the other hand, the figure also shows the relative error for the initial tensor constructed using the character expansions in Eq.~\eqref{eqn:tensor_ch}.
In this case, the allowed value of the bond dimension $D_{\rm c}$ is severely restricted as explained in Appendix \ref{sec:TN_c} and we follow the truncation order adopted in the previous study \cite{PhysRevD.105.054507}.
The resulting error is of order $O(10^{-3})$ for the parameter sets we investigated.
In conclusion, our new initial tensor, constructed using the quadrature method together with the compression, is shown to be better than that based on the character expansion.
Therefore, in the subsequent numerical calculations, we exclusively employ the new initial tensor.

\section{Numerical results}
\label{sec:result}
To investigate the phase structure, we compute various physical quantities by coarse-graining the new initial tensor defined in the previous section.
In the following calculations, the parameters for the quadrature-based initial tensor are set to $m=6$, $N_{\rm A}=120$, and $D_{\rm c}=128$.
For the coarse-graining procedure, we employ the bond-weighted tensor renormalization group algorithm (bwTRG) \cite{PhysRevB.105.L060402} with the hyperparameter $k=-\frac{1}{2}$.
Unless otherwise stated, the bond dimension for the coarse-graining is taken to be the same as that used in the initial tensor.

\subsection{Free energy}

First, let us see the free energy density defined as
\begin{equation}
    f=-\frac{\ln Z}{V}\approx-\frac{\ln \sum_{ij}T^{(n)}_{ijij}}{V},
\label{eqn:f}
\end{equation}
where $T^{(n)}$ denotes the tensor obtained after $n$ renormalization steps, and the lattice volume is given by $V=L^2=4^n$. 
It should be noted that one renormalization step consists of two coarse-graining procedures of the bwTRG algorithm.

\begin{figure}[tb]
    \centering
    \includegraphics[width=0.6\columnwidth]{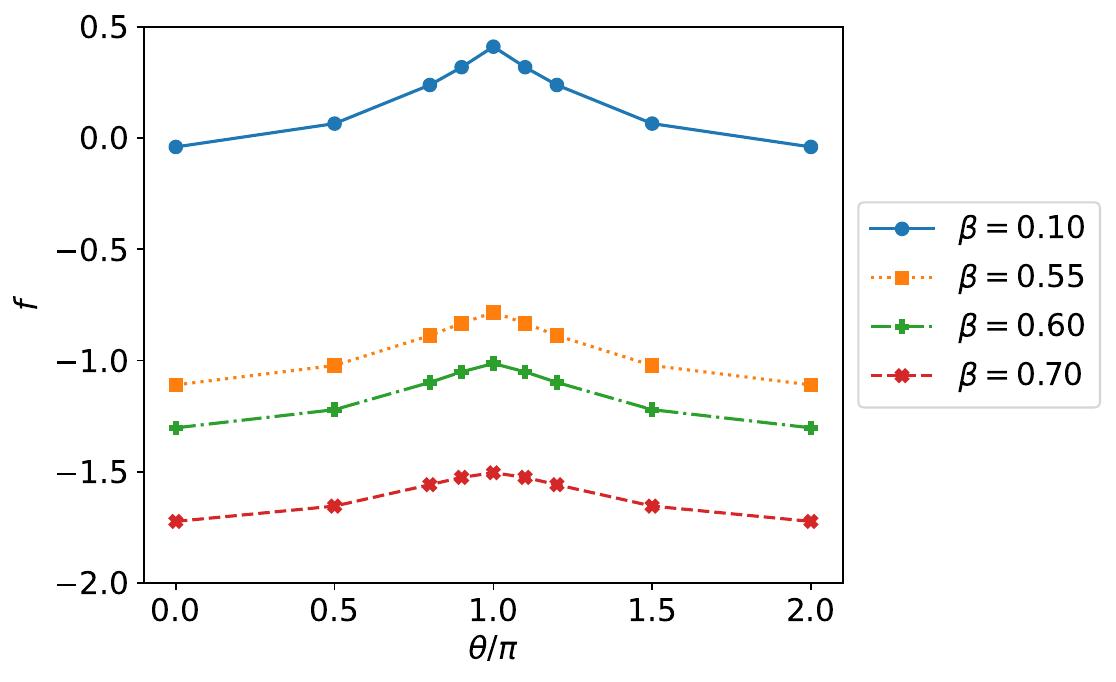}
   \caption{The free energy density in Eq.~\eqref{eqn:f} as a function of $\theta$ on $V=4^5$ for several values of $\beta=0.1$ -- $0.7$. 
	The shape cusp structure at small $\beta$ becomes smoother for large $\beta$.
   }
   \label{fe-theta}
\end{figure}

Figure~\ref{fe-theta} shows the free energy density as a function $\theta$.
The free energy is symmetric about $\theta=\pi$.  
In particular, at $\beta = 0.1$, the free energy exhibits a clear cusp at $\theta = \pi$, and this
indicates a first-order phase transition and it is consistent with the strong coupling analysis \cite{PhysRevD.55.3966}.
On the other hand, the cusp becomes smoother as $\beta$ increases, and it is expected to turn out to be the second-order phase transition at some value of $\beta$.
It is, however, difficult to determine the transition point from the free energy itself.

\subsection{Degeneracy of ground state}
In order to quantitatively identify the nature of transition especially for the first-order phase transition, we employ the ground state degeneracy, which can be probed by $X$ \cite{PhysRevB.80.155131},
\begin{equation}
    X=\frac{(\sum_kM_{kk})^2}{\sum_{ij}M_{ij}M_{ji}},
\label{eqn:X}
\end{equation}
where $M$ denotes the transfer matrix of the system.
The transfer matrix is constructed from the coarse-grained tensor $T^{(n)}$ by contracting along the spatial direction 
\begin{equation}
    M_{ij}=\sum_{k}T_{ikjk}^{(n)}.
\end{equation}
For sufficiently large lattice size, one obtains $X=2$ when the ground state is doubly degenerate 
whereas in the absence of degeneracy one finds $X=1$.
When a $Z_2$ symmetry (in our case, CP symmetry) is spontaneously broken, the ground state is doubly degenerate.
Thus, $X=2$ serves as a reliable quantitative indicator of a first-order phase transition for this case.

\begin{figure}[tbh]
    \centering
    \includegraphics[width=0.7\columnwidth]{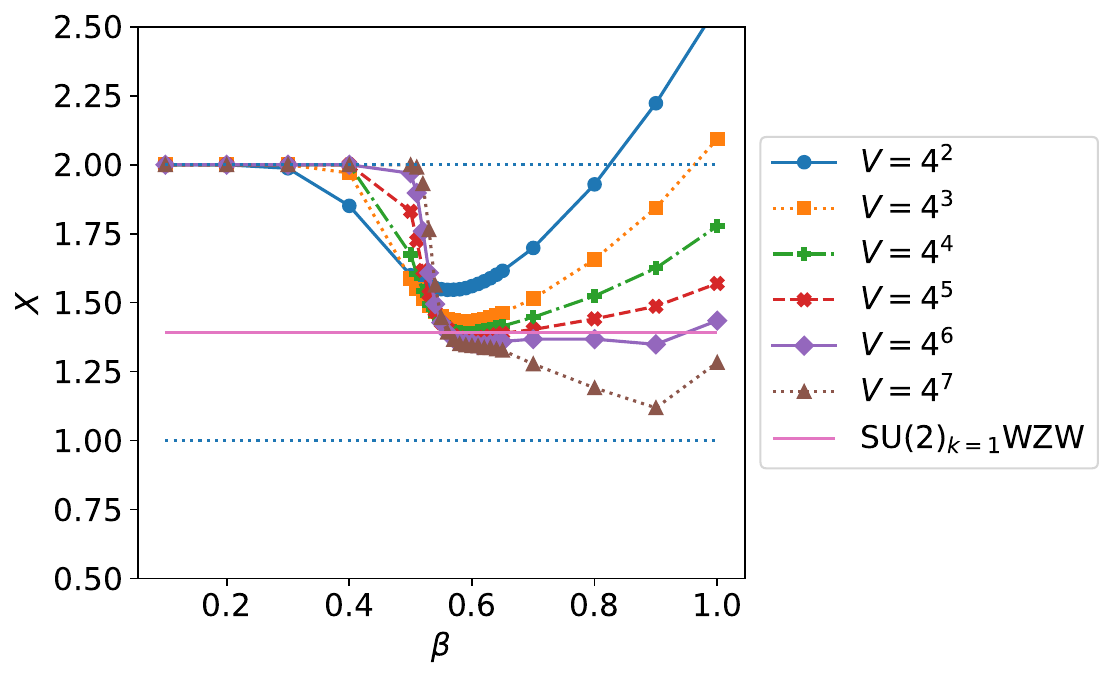}
   \caption{Ground-state degeneracy $X$ in Eq.~\eqref{eqn:X} as a function of $\beta$ at $\theta=\pi$ on $V=4^2$ -- $4^7$.
   The pink line represents the value for SU(2)$_{k=1}$ WZW model in Eq. \eqref{x-wzw}.}
   \label{fig:Xtpi}
\end{figure}

\begin{figure}[tbh]
    \centering
\begin{minipage}{0.45\columnwidth}
    \centering
    \includegraphics[width=0.9\columnwidth]{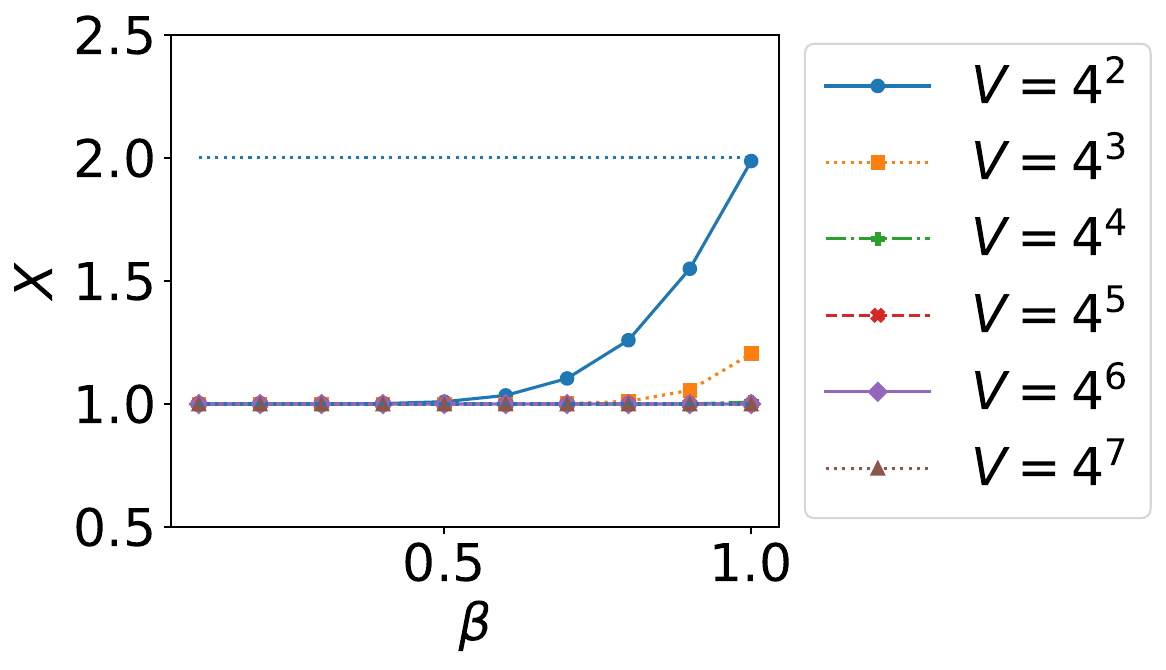}
    \subcaption{$\theta=0$}
\end{minipage}
\begin{minipage}{0.45\columnwidth}
    \centering
    \includegraphics[width=0.9\columnwidth]{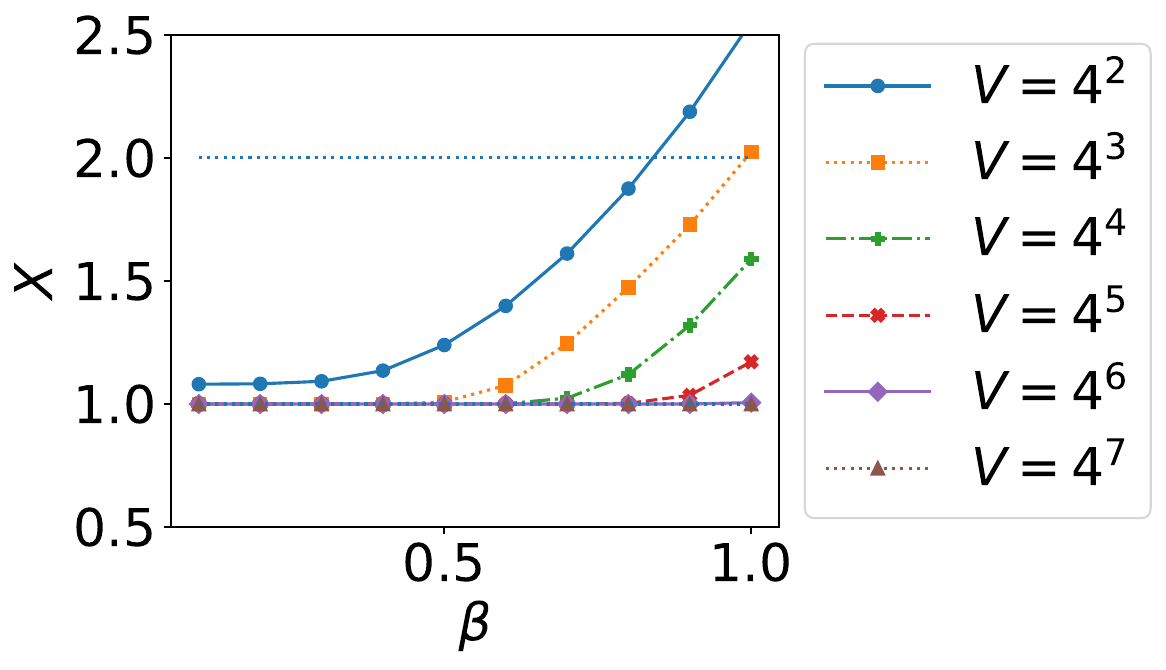}
    \subcaption{$\theta=0.9\pi$}
\end{minipage}
   \caption{The $\beta$ dependence of $X$ for $\theta\neq\pi$ on $V=4^2$ -- $4^7$.
   The results for $\theta=0$ and $\theta=0.9\pi$ are shown in panels (a) and (b), respectively.}
   \label{fig:Xtnpi}
\end{figure}

Figure~\ref{fig:Xtpi} shows $X$ as a function of $\beta$ at $\theta = \pi$.
For $\beta<0.5$, we find that $X = 2$ persists even for sufficiently large volumes, 
thereby, confirming the presence of the first-order phase transition in the low $\beta$ region.
In contrast, for $\beta>0.5$, $X$ takes non-integer values, and thus the presence or absence of ground-state degeneracy cannot be unambiguously determined.
When the system is tuned to its critical point, the value of $X$ can be represented using the partition function of the corresponding CFT.
The modular invariant partition function for SU(2)$_{k=1}$ WZW model on torus \cite{DiFrancesco:1997nk,Ginsparg:1988ui} is known to be
\begin{equation}
    Z(\tau)=\left|\frac{1}{\eta}\sum_{n\in\mathbb{Z}}q^{n^2}\right|^2+\left|\frac{1}{\eta}\sum_{n\in\mathbb{Z}}q^{(n+\frac{1}{2})^2}\right|^2,
\hspace{5mm}
    q=e^{2\pi i \tau}
    \end{equation}
where $\tau\in \mathbb{C}$ is the modular parameter 
and $\eta$ is the Dedekind $\eta$ function defined by 
\begin{equation}
    \eta(\tau)=q^{\frac{1}{24}}\prod_{n=1}^{\infty}(1-q^n).
\end{equation}
In terms of the partition function, the value of $X$ for the SU(2)$_{k=1}$ WZW model is given by
\begin{equation}
    X=\frac{Z(\tau=i)^2}{Z(\tau=2i)}=1.39324280(1). 
    \label{x-wzw}
\end{equation}
This value is shown as the pink line in Fig.~\ref{fig:Xtpi}, and the numerical data for $X$ approaches this line as the volume is increased for $\beta = 0.55\text{ -- }0.6$.
This behavior suggests that the critical point associated with the SU(2)$_{k=1}$ WZW model lies within this range of $\beta$.
To determine the critical point more accurately, we will perform a more detailed analysis using CFT information in the next subsection.

On the other hand, Fig.~\ref{fig:Xtnpi} displays $X$ for the case $\theta \neq \pi$, where we consistently obtain $X = 1$ for all $\beta$ in sufficiently large volumes. 
This result clearly indicates that no first-order phase transition occurs away from $\theta=\pi$ within the $\beta$ region we investigated.

\newpage
\subsection{Analysis based on CFT data}

As seen in the previous subsections, 
the analyses of the free energy and $X$ establish that the system undergoes a first-order phase transition in the low $\beta$ region at $\theta=\pi$.
On the other hand, the remaining task is to determine the value of $\beta$ at which the first-order phase transition terminates and to identify the order of the transition beyond that point.
In the large $\beta$ region, the weakening of the transition strength suggests the occurrence of a second-order phase transition, which is also consistent with the Haldane conjecture. 
When dealing with systems that are expected to exhibit a criticality in two dimensions, it is highly beneficial to make use of the knowledge of conformal field theory (CFT).
Criticality is classified by its universality class, with a corresponding CFT associated to each class.
Furthermore, CFT itself is characterized by fundamental quantities such as the central charge and scaling dimensions.
In fact, these quantities can be computed by the TRG methods \cite{PhysRevB.80.155131}.
Once an invariant tensor is obtained after several iterations of the coarse-graining, the transfer matrix constructed from the invariant tensor can be regarded as the transfer matrix of the corresponding CFT. 
The eigenvalues $\lambda_i$ of the transfer matrix are then related to the central charge $c$ and scaling dimensions $x_i$ ($i = 1, 2, 3, \dots$) as follows,
\begin{equation}
    c=\frac{6}{\pi}\log(\lambda_0),
    \hspace{10mm}
    x_i=\frac{1}{2\pi}\log\left(\frac{\lambda_0}{\lambda_i}\right).
\end{equation}
Strictly speaking, these quantities receive corrections from irrelevant operators \cite{CARDY1986186,JLCardy_1986}.
However, since the leading term of the correction to the central charge is $O(\frac{1}{(\log L)^3})$, it can be neglected for sufficiently large volumes.
The correction to the scaling dimension will be discussed later.

\begin{figure}[t]
    \centering
    \includegraphics[width=0.7\columnwidth]{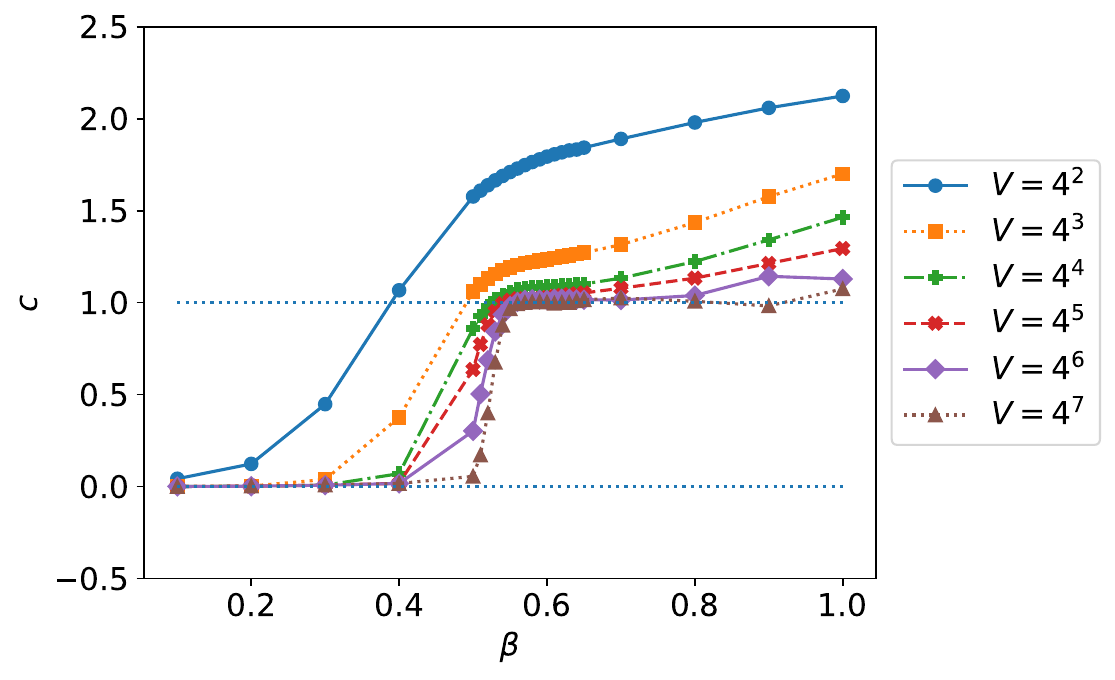}
   \caption{The central charge $c$ as a function of $\beta$ at $\theta=\pi$ on $V=4^2$ -- $4^7$.}
   \label{fig:cp1_th1_beta_cc_tpi}
\end{figure}

\begin{figure}[t]
    \centering
\begin{minipage}{0.45\columnwidth}
    \centering
    \includegraphics[width=0.9\columnwidth]{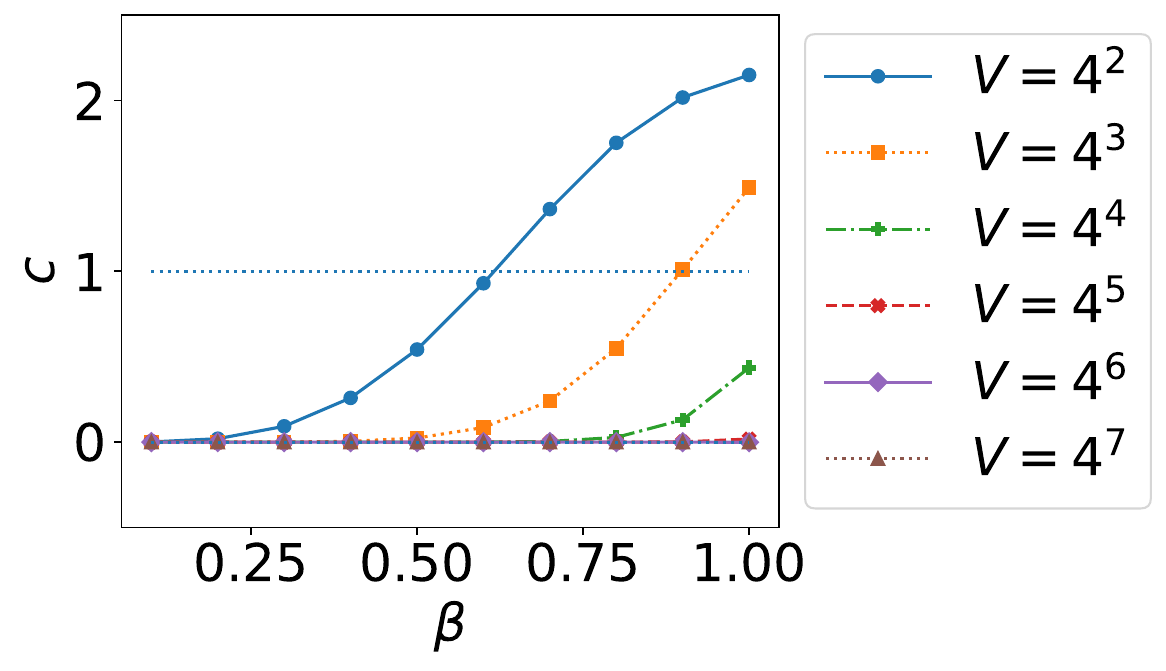}
    \subcaption{$\theta=0$}
\end{minipage}
\begin{minipage}{0.45\columnwidth}
    \centering
    \includegraphics[width=0.9\columnwidth]{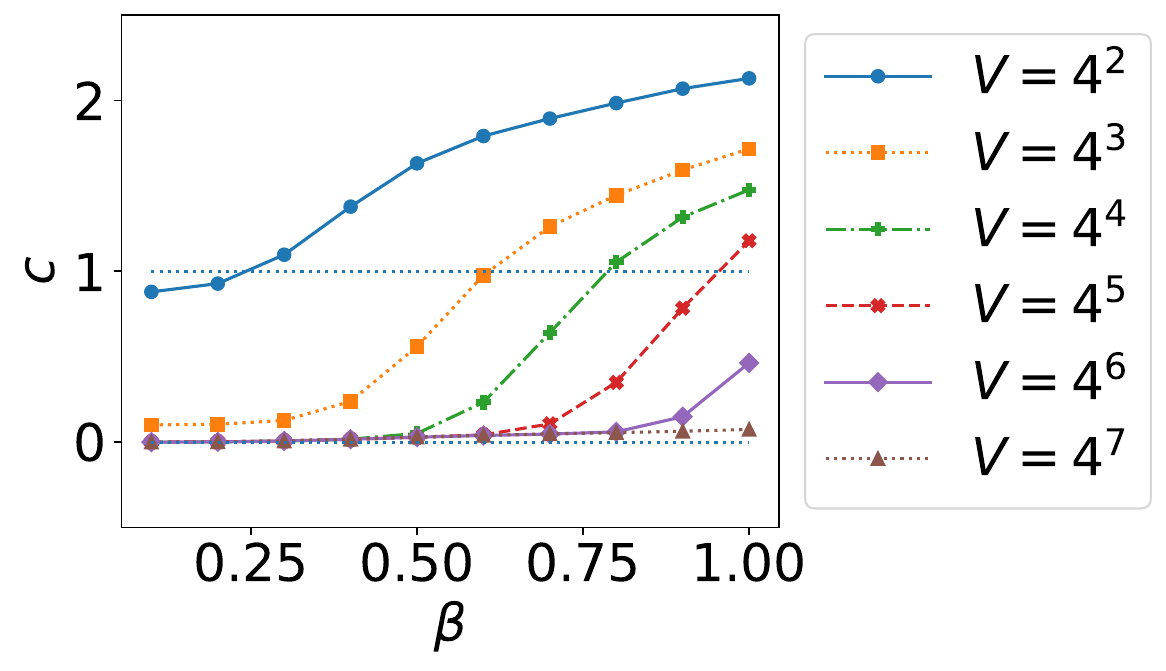}
    \subcaption{$\theta=0.9\pi$}
\end{minipage}
   \caption{The central charge for $\theta\neq\pi$ on $V=4^2$ -- $4^7$.
   The results for $\theta=0$ and $\theta=0.9\pi$ are shown in panels (a) and (b), respectively.}
   \label{fig:cp1_th1_beta_cc_tnpi}
\end{figure}

Numerical results of the central charge at $\theta=\pi$ are shown in Fig.~\ref{fig:cp1_th1_beta_cc_tpi}.
The central charge remains zero in the low-$\beta$ region, but abruptly jumps to unity as $\beta$ increases.
Moreover, this tendency becomes shaper as the lattice volume increases.
Therefore, we can see that the region $\beta \gtrsim 0.55$ at $\theta=\pi$ is a critical regime, and its universality class is that of free bosons.
On the other hand, the central charge for $\theta\neq\pi$ is shown in Fig.~\ref{fig:cp1_th1_beta_cc_tnpi}, 
where it approaches zero with increasing system volume across the entire $\beta$ region investigated.
Therefore, in this region, we conclude not only the absence of a first-order phase transition but also that no second-order phase transition occurs.

The phase diagram inferred from the analyses of the ground-state degeneracy $X$ and the central charge is basically identical to Fig.~\ref{fig:phase_structure_CP1_pre}.
Specifically, at $\theta=\pi$, a first-order phase transition occurs in the low-$\beta$ region, whereas in the high-$\beta$ region the system undergoes a second-order phase transition.
The remaining task is to identify the point at which the transition changes from first-order to second-order.
Before discussing this identification, it is worth noting that the phase diagram containing this kind of critical line resembles that of the two-dimensional XY model, which exhibits the BKT transition \cite{JMKosterlitz_1973,JMKosterlitz_1974}.
The universality class of the BKT transition points, at which the critical line originates, corresponds to the SU(2)$_{k=1}$ WZW model. 
This universality class is identical to that of the O(3) nonlinear $\sigma$ model at $\theta=\pi$, as predicted by
the Haldane conjecture \cite{PhysRevB.36.5291,PhysRevLett.66.2429}
and more recently confirmed by numerical calculations
\cite{PhysRevLett.75.4524,PhysRevD.77.056008,PhysRevD.86.096009}.
According to the analytic consideration \cite{PhysRevLett.55.1355,JLCardy_1986},
around the BKT transition point, logarithmic corrections to the scaling dimension arise from a marginally irrelevant perturbation.
These corrections render the identification of the transition point by the conventional finite-size scaling methods particularly difficult \cite{PhysRevB.48.16814}. 
A solution to this problem was proposed in Ref.~\cite{KNomura_1994}, which relies on level spectroscopy based on CFT data, and it has recently been further refined by incorporating tensor renormalization group (TRG) techniques \cite{PhysRevB.104.165132}.
In fact, this method has determined the BKT transition point with remarkably high accuracy.

\begin{figure}[t]
    \centering
    \includegraphics[width=0.8\columnwidth]{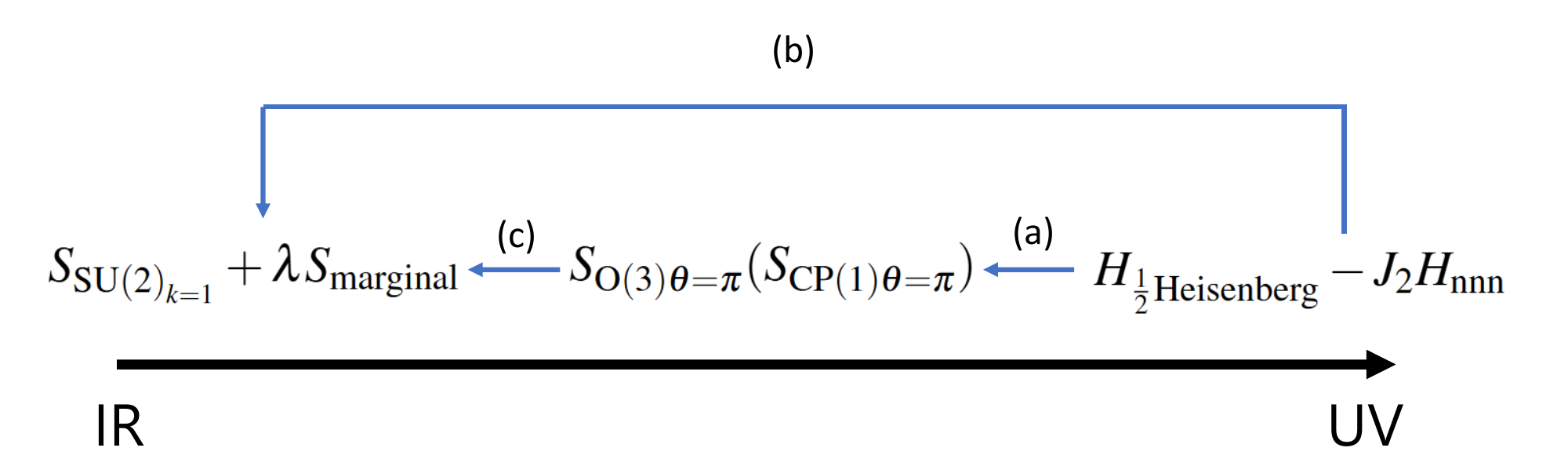}
   \caption{The relation between the Heisenberg model, O(3) nonlinear $\sigma$ model, and the SU(2)${}_{k=1}$ WZW model.
   The arrow (a) means that the O(3) model at $\theta=\pi$ is the low-energy effective field theory of a spin-$\frac{1}{2}$ Heisenberg model with an additional next-nearest-neighbor interaction~\cite{Affleck:1984ar,PhysRevB.36.5291}.
   The arrow (b) represents that the Heisenberg model corresponds to the SU(2)${}_{k=1}$ WZW model with an additional marginal operator in the low-energy limit, which can be shown by using the bosonization procedure~\cite{Affleck:1985wb, PhysRevB.36.5291}.
   The arrow (c) is predicted from combining (a) and (b), and it states that O(3) nonlinear $\sigma$ model at $\theta=\pi$ also corresponds to the SU(2)${}_{k=1}$ WZW model with an additional marginal operator in the low-energy limit.
   }
   \label{eff-th}
\end{figure}

In the present work, we also adopt the level spectroscopy method to investigate the phase structure of the lattice CP(1) model,
which is equivalent to O(3) nonlinear $\sigma$ model in the continuum limit.
However, note that while the O(3) model possesses a global SO(3) symmetry, the XY model does not, and consequently
the perturbations allowed around the corresponding CFT differ between the two cases.
Therefore, for a while, let us briefly discuss the perturbation for the case of the O(3) model.
The form of the perturbation is predicted by combining the relation between antiferromagnets and O(3) known as the Haldane conjecture with the application of bosonization to antiferromagnets.
Figure \ref{eff-th} schematically represents this prediction.
The arrow (a) in Fig.~\ref{eff-th} means that the O(3) nonlinear $\sigma$ model at $\theta=\pi$ is the low-energy effective field theory of a spin-$\frac{1}{2}$ Heisenberg model with an additional next-nearest-neighbor interaction, in the regime where the next-nearest-neighbor coupling is sufficiently large~\cite{Affleck:1984ar,PhysRevB.36.5291}.
In this correspondence, the coupling constant $J_2$ of the next-nearest-neighbor interaction in the Heisenberg model maps onto the inverse coupling $\beta$ of the O(3) model.
On the other hand, in such large $J_2$ case, the Heisenberg model corresponds to the SU(2)${}_{k=1}$ WZW model with an additional marginal operator in the low-energy limit
as shown in the arrow (b) in Fig.~\ref{eff-th}. 
This correspondence is revealed by non-Abelian bosonization analyses~\cite{Affleck:1985wb, PhysRevB.36.5291}, and
note that $J_2$ maps onto $\lambda$, the coupling constant associated with the marginal operator.
From the above arguments (a) and (b), the relation indicated by the arrow (c) in Fig.~\ref{eff-th} can be predicted, i.e.,
\begin{equation}
    S_{\mathrm{O(3)}\theta=\pi}(S_{\mathrm{CP(1)}\theta=\pi} )    
    \sim S_{\mathrm{SU(2)}_{k=1}}+a(\beta-\beta_{\rm c})S_{\rm marginal},
\end{equation}
where the marginal coupling constant $\lambda$ 
are related with $\beta$ in the O(3) model as $\lambda=a(\beta-\beta_{\rm c})$ with a proportional constant $a$.
In the equation above, the symbol $\sim$ indicates that irrelevant contributions are ignored (these contributions will be discussed later).
Here, $S_{\mathrm{SU(2)}_{k=1}}$ has the chiral symmetry $\mathrm{SU(2)}_{\rm L}\times\mathrm{SU(2)}_{\rm R}$, and it is known that the lowest scaling dimension is $\frac{1}{2}$, which comes with a fourfold degeneracy.
On the other hand, $S_{\rm marginal}$ preserves only the diagonal subgroup of the chiral symmetry,\footnote{In Appendix \ref{sec:symmetry}, we investigate to what extent the SO(3) symmetry and the CP symmetry at $\theta=\pi$ are maintained in our numerical calculation.}
SO(3), and consequently the fourfold degeneracy is lifted into a singlet and a triplet once $\beta$ is shifted away from $\beta_{\rm c}$.
This is a crucial distinction between the XY model and the O(3) model.
In the end, perturbation calculations for the SU(2)${}_{k=1}$ WZW model \cite{IAffleck_1989,OKAMOTO1992433} yield the scaling dimensions of the singlet $x_{\rm s}$ and triplet $x_{\rm t}$ as
\begin{equation}
    x_{\rm s}=\frac{1}{2}+\frac{3}{4}\frac{2}{2\log(L)+\frac{1}{a(\beta-\beta_{\rm c})}}\underset{\beta\to\beta_{\rm c}}{\approx}\frac{1}{2}+\frac{3a}{2}(\beta-\beta_{\rm c}),
\end{equation}
\begin{equation}
    x_{\rm t}=\frac{1}{2}-\frac{1}{4}\frac{2}{2\log(L)+\frac{1}{a(\beta-\beta_{\rm c})}}\underset{\beta\to\beta_{\rm c}}{\approx}\frac{1}{2}-\frac{a}{2}(\beta-\beta_{\rm c}).\end{equation}
It is evident that the fourfold degeneracy is restored at $\beta=\beta_{\rm c}$, and the deviation is controlled by the common constant $a$.

\begin{figure}[t]
    \centering
    \includegraphics[width=0.7\columnwidth]{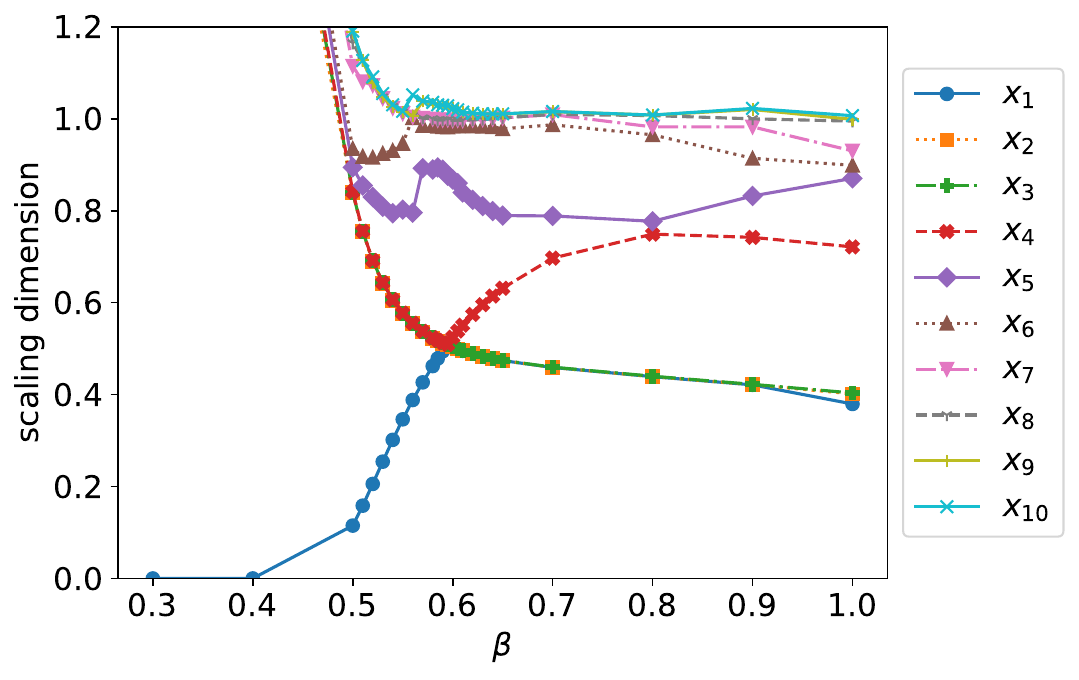}
    \caption{Scaling dimensions $x_i$ ($i=1,2,\ldots,10$) as a function of $\beta$ on $V=4^5$ at $\theta=\pi$. The four lowest scaling dimensions exhibit a crossing behavior around $\beta\approx0.6$.}
    \label{fig:sdall}
\end{figure}
\begin{figure}[ptb]
    \centering
    \includegraphics[width=0.7\columnwidth]{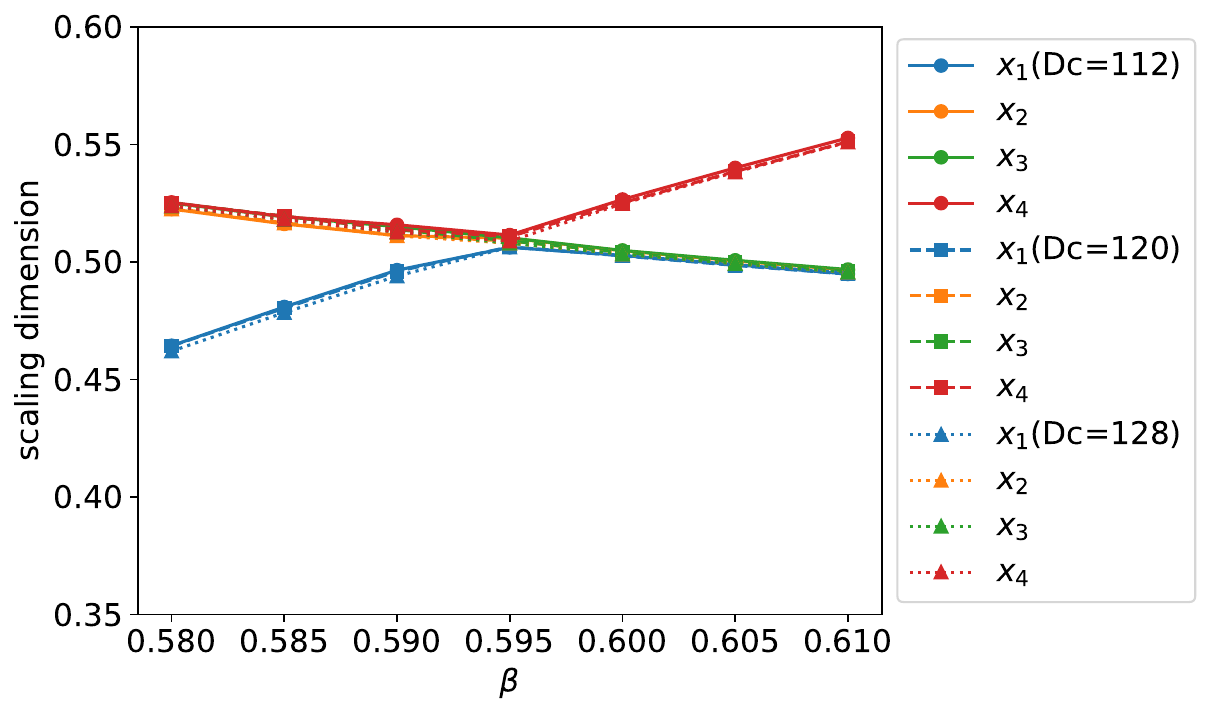}
   \caption{The four lowest scaling dimensions, $x_1$, $x_2$, $x_3$, and $x_4$, near the transition point on $V=4^5$, computed for three bond dimensions, $D_{\rm c}=112, 120$, and $128$. In this range of $\beta$, the scaling dimensions exhibit linear behavior, and the dependence of $D_{\rm c}$ is rather small.}
   \label{fig:bdep}
\end{figure}

\begin{figure}[ptb]
    \centering
    \includegraphics[width=0.7\columnwidth]{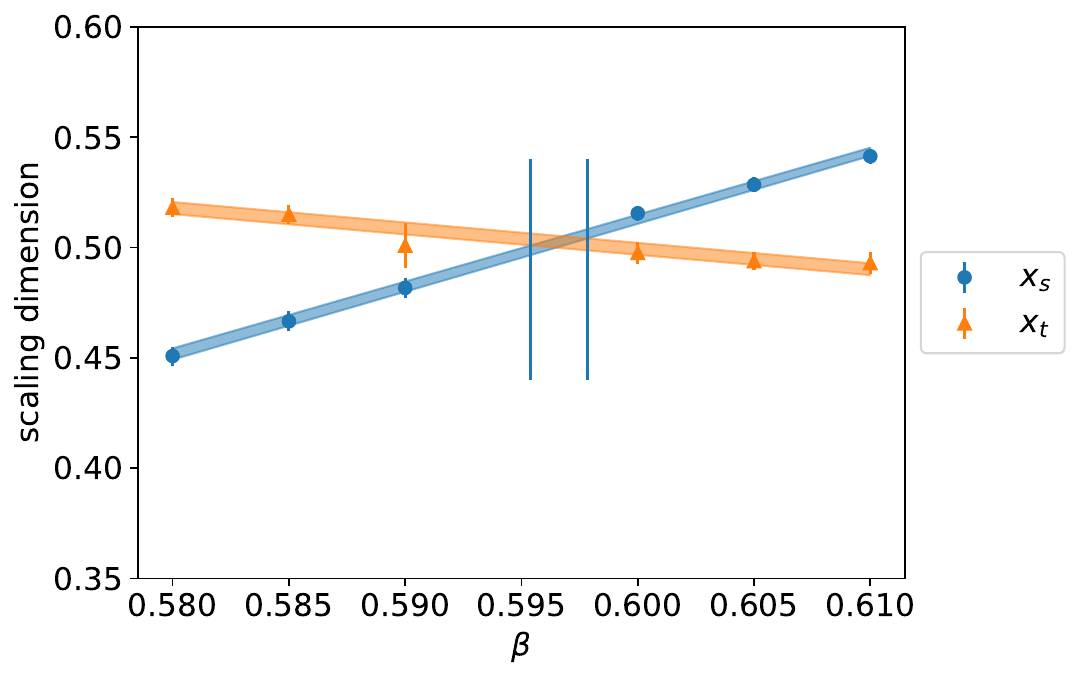}
   \caption{$\beta$-interpolation of the scaling dimension of the single and triplet on $V=4^5$.
   In the interpolation, the data at $\beta=0.595$ are omitted since it is difficult to distinguish between the singlet or triplet sectors.
   The vertical lines show the boundaries of the overlap region between the singlet and triplet interpolation bands.}
   \label{fig:bint}
\end{figure}

\begin{figure}[ptb]
    \centering
    \includegraphics[width=0.6\columnwidth]{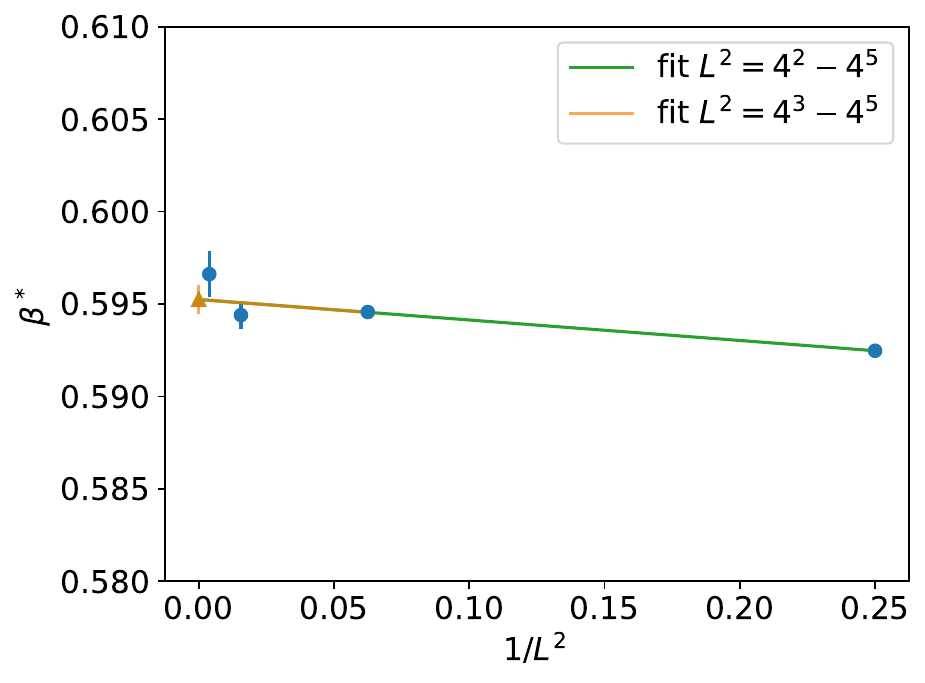}
   \caption{Infinite-volume extrapolation of $\beta^\ast$. The fourfold degeneracy point $\beta_{\rm c}$ is determined by using Eq.~\eqref{Vex}.}
   \label{fig:betac}
\end{figure}

Figure \ref{fig:sdall} shows the ten lowest scaling dimensions obtained using our method. 
As shown in the figure, the four lowest scaling dimensions split into a singlet and a triplet, and the transition point can be roughly estimated as $\beta_{\rm c}\approx 0.6$ where the fourfold degeneracy is observed.
Thus, our numerical results are consistent with the prediction of the arrow (c) in Fig.~\ref{eff-th}.
Figure~\ref{fig:bdep} shows that, in the vicinity of the transition point, the four lowest scaling dimensions vary nearly linear dependence on $\beta$.
In the figure, the results for three values of $D_{\rm c}$ are also presented, showing that the bond dimension effects are small.
Here, note that the bond dimension for the initial tensor is fixed to $D_{\rm c}=128$, while the bond dimension for the bwTRG is only changed in the range $112$ -- $128$ to see the truncation error caused by the coarse-graining procedure.
We perform the $D_{\rm c}$-extrapolation of each scaling dimension using both linear and the quadratic fits in $1/D_{\rm c}$, and then the error is taken as the difference between the two fits.\footnote{
For the triplet sector, the three scaling dimensions are slightly lifted due to finite-$D_{\rm c}$ effects. 
Thus, we first extrapolate the three scaling dimensions to the $D_{\rm c}\to\infty$ limit individually, using the two fitting forms described in the main text. Among the six extrapolated values, we take the average of the maximum and minimum as the central value of the triplet, and define their difference as the error.}
The error obtained in the extrapolation is shown in each data in Fig.~\ref{fig:bint}.
Then, adopting the linear functional form in $\beta$, we interpolate the scaling dimensions for the singlet and triplet sectors separately.\footnote{In this linear interpolation procedure, we perform independent interpolations for the singlet and triplet scaling dimensions, each with its own coefficient in $\beta$.}
The resulting interpolations are shown as bands in Fig.~\ref{fig:bint}.
From the overlap region of the singlet and triplet interpolation bands, 
we estimate the crossing point $\beta^\ast$ and its uncertainty:
the midpoint of the overlap region is taken as the central value of $\beta^\ast$, while its half-width is assigned as the error.
The resulting $\beta^\ast$ for several volumes $V=4^2$ -- $4^5$ are shown in Fig.~\ref{fig:betac}.
Small volume dependence is observed, which can be attributed to the effects of irrelevant operators that have been neglected in the discussion so far.
According to CFT analyses \cite{IAffleck_1989,OKAMOTO1992433},
the leading contribution to the scaling dimension from the irrelevant terms is proportional to $1/L^2$,
\begin{equation}
    x_{\rm s}=\frac{1}{2}+\frac{3a}{2}(\beta-\beta_{\rm c})+\frac{C_{\rm s}}{L^2},
\end{equation}
\begin{equation}
    x_{\rm t}=\frac{1}{2}-\frac{a}{2}(\beta-\beta_{\rm c})+\frac{C_{\rm t}}{L^2}.
\end{equation}
Using these expressions, the crossing point $\beta^\ast$ is determined from the condition $x_{\rm s}=x_{\rm t}$, yeilding
\begin{equation}
    \beta^*=\beta_{\rm c}+\frac{C_{\rm t}-C_{\rm s}}{2aL^2}. 
    \label{Vex}
\end{equation}
Therefore, the transition point $\beta_{\rm c}$ is obtained by extrapolating $\beta^\ast$ to the infinite-volume limit. 
See Fig.~\ref{fig:betac} for the extrapolation.

\begin{figure}[ptb]
    \centering
    \includegraphics[width=0.7\columnwidth]{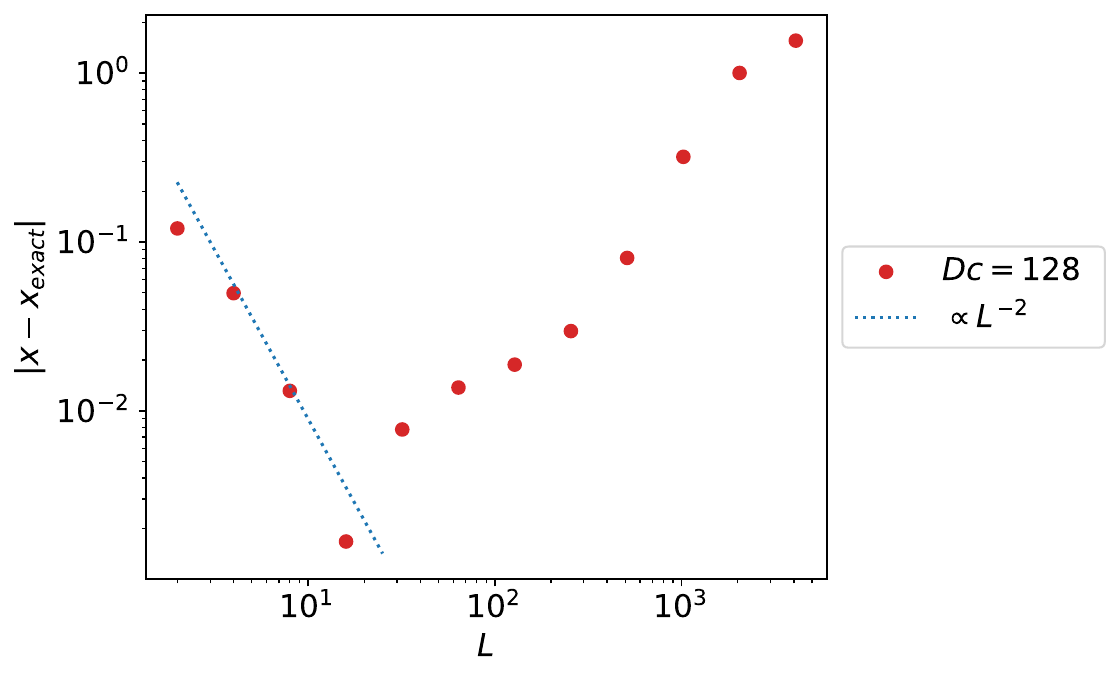}
   \caption{The deviation $|x-x_{\rm exact}|$ as a function of $L$ at $\beta=0.595$ with $D_{\rm c}=128$. The $1/L^2$ scaling arising from the irrelevant contribution is seen in the small $L\le16$, while the deviation starts to increase from $L=32$ which is attributed from the finite bond dimension effects.}
   \label{fig:Dcscale}
\end{figure}

Finally, we comment on the fitting range in the infinite-volume extrapolation.
In the extrapolation, the largest volume is chosen as $V=4^5$, and the reason for this choice is explained below.
We first consider the average of the four lowest scaling dimensions, $x=\frac{1}{4}\sum_{i=1}^4x_i\approx\frac{x_{\rm s}+3x_{\rm t}}{4}$, for which the logarithmic correction is expected to cancel out.
Figure \ref{fig:Dcscale} shows a deviation, $|x-x_{\rm exact}|$, from the exact value $x_{\rm exact}=1/2$ as a function of $L$ at $\beta=0.595$,
which is the point closest to the crossing.
The deviation roughly follows a $1/L^2$ scaling for smaller volumes up to $L\sim2^4=16$, which is the expected behavior arising from the irrelevant operator.
However, the deviation starts to increase once $L$ exceeds $16$, and this behavior is attributed from systematic errors caused by the finite bond dimension $D_{\rm c}$, namely emergent relevant contributions  \cite{Ueda:2023smj}.
We confirm that a similar behavior is observed for other values of $\beta=0.58$ -- $0.61$.
Hence, to avoid the finite bond dimension effects, we use only the data that follow the $1/L^2$ scaling and for which the deviation is below $10^{-2}$,
that is, we use $V=4^2$ -- $4^5$ for the analysis.
Figure~\ref{fig:betac} presents the infinite-volume extrapolation using data for volumes up to $V\le4^5$.
We perform the extrapolation using two fitting ranges, depending on whether the smallest volume is included or not, and find that the difference between the results is negligible.
Finally, the transition point is determined as $\beta_{\rm c}=0.5952(8)$.

\section{Conclusion}
\label{sec:conclusion}
 We have investigated the phase structure of the two-dimensional lattice CP(1) model with $\theta$ term using the bond-weighted TRG, and observed a critical region along the $\theta = \pi$ line.
In this study, we made two main improvements over previous studies: (i) a new initial tensor constructed with the quadrature method, and (ii) a CFT-based analysis method.
The new initial tensor was shown to provide higher accuracy than the conventional one.
Furthermore, by utilizing the central charge and the scaling dimensions of CFT, we identified the critical region and determined its onset to be
$\beta_{\rm c} = 0.5952(8)$ at $\theta=\pi$ where the SU(2)$_{k=1}$ WZW model is realized.  
The universality class of this critical point is consistent with the prediction of the Haldane conjecture.

Having clarified the phase structure including the $\theta$ term, we are now in a position to investigate the continuum CP(1) +$\theta$ model non-perturbatively.
As a next step, we plan to compute the renormalized coupling constant \cite{LUSCHER1991221} and the string tension \cite{PhysRevLett.53.637}, and then take the continuum limit to reveal the impact of the $\theta$ term on these physical quantities.
Moreover, the phase structure of the CP(2) model is also non-trivial and of interest.
This model may be analyzed using the large-$N$ expansion, and we intend to examine whether the results are consistent with those reported in \cite{DAdda:1978vbw,Witten:1978bc}.  
We expect that these analyses will lead to a deeper understanding of the role of the $\theta$ term.

\section*{Acknowledgement}
We would like to thank Yoshinobu Kuramashi and Yuya Tanizaki for the useful discussions.
This work was partially supported by JSPS KAKENHI Grant Number
21K03531,	
22H05251,	
and 25K07280.	
This work was supported by JST SPRING, Grant Number JPMJSP2135.

\appendix
\section{Tensor network representation using character expansion}
\label{ch}
\subsection{Character-like expansion}
In this subsection, we consider CP($N-1$) with arbitrary $N$ case, namely the complex scalar field is now $z(x)\in\mathbb{C}^N$ and is subject to $|z(x)|^2=\sum_{i=1}^N |z_i(x)|^2=1$.
 Equation \eqref{eq:H} may be expanded in terms of orthogonal functions $f_{lm}$ as follows \cite{PhysRevD.55.3966},
\begin{equation}
    H_{vwA}=e^{N\beta(v^\dagger we^{iA}+w^{\dagger} ve^{-iA})}=Z_0(\beta)\sum_{l,m\in\mathbb{Z}}d_{lm}h_{lm}(\beta)e^{i(m-l)A}f_{lm}(v,w).
    \label{chl}
\end{equation}
This expansion is called a character-like expansion.
We may define the orthogonal function $f_{lm}(v,w)\in\mathbb{C}$ as follows,
\begin{equation}
    f_{lm}(v,w)=\sum_{n=0}^{\min(l,m)}F_{n}^{lm}(v\cdot w^*)^{l-n}(v^*\cdot w)^{m-n}.
    \label{Flmndef}
\end{equation}
The coefficient $F_n^{lm}\in\mathbb{R}$ is defined so that $f_{lm}$ satisfies the orthogonality relation,
\begin{equation}
\int dz f_{lm}(v,z)f_{l^\prime m^\prime}^\ast(w,z)
=
\frac{1}{d_{lm}}
\delta_{ll^\prime}
\delta_{mm^\prime}
f_{lm}(v,w),
\label{eq:gen_orthogonal_relation_f}
\end{equation}
where $\int dz$ is an integral measure for the complex scalar field,
and the dimension $d_{lm}$ is given by
\begin{equation}
    d_{lm}=f_{lm}(v,v)=\sum_{n=0}^{\min(l,m)}F_{n}^{lm}.
    \label{dlmdef}
\end{equation}
In particular, for $v=w$, the standard orthogonality relation holds
\begin{equation}
    \int dz f_{lm}(v,z)f_{l^\prime m^\prime}^\ast(v,z)
    =
    \delta_{ll^\prime}
    \delta_{mm^\prime}.
    \label{eq:orthogonal_relation_f}
\end{equation}
Note that this holds for any $v$.
In eq.(\ref{chl}), $h_{lm}\in\mathbb{R}$ is a coefficient of the character-like expansion and depends on $\beta$.
A normalization factor $Z_0(\beta)\in\mathbb{R}$ is determined such that $h_{00}=1$.

From Eq.~\eqref{Flmndef} and \eqref{eq:orthogonal_relation_f}, 
it can be shown that $f_{lm}$ obeys a three-term recurrence relation,
\begin{equation}
    f_{lm}(v,w)=\alpha_{lm}\left(|v\cdot w^*|^2+\beta_{lm}\right)f_{l-1,m-1}(v,w)+\gamma_{lm}f_{l-2,m-2}(v,w),
    \label{eq:3rec}
\end{equation}
with coefficients
\begin{eqnarray}
\alpha_{lm}&=&\frac{F_{0}^{lm}}{F_{0}^{l-1,m-1}},
\\
\beta_{lm}&=&-\int dv^\prime|v^\prime\cdot w^{\prime*}|^2f_{l-1,m-1}(v^\prime,w^\prime)f_{l-1,m-1}^*(v^\prime,w^\prime),
\\
\gamma_{lm}&=&-\frac{F_{0}^{lm}F_{0}^{l-2,m-2}}{(F_{0}^{l-1,m-1})^2}.
\end{eqnarray}
This recurrence relation, together with Eq.~(\ref{Flmndef}), provides a recurrence relation for $F^{lm}_{n}$.
One can verify that the following expression satisfies the relation,
\begin{equation}
    F_{n}^{lm}=\sqrt{\frac{(N-1+l+m)l!m!}{(N-2+l)!(N-2+m)!(N-1)}}(-1)^n\frac{(N-2+l+m-n)!}{(l-n)!(m-n)!n!}.
\label{eqn:Fnlm}
\end{equation}
Substituting Eq.~\eqref{eqn:Fnlm} into Eq.~\eqref{Flmndef}, one can check that Eq.~\eqref{eq:orthogonal_relation_f} holds.
Using the closed-form of $F^{lm}_n$ above,
$d_{lm}$, $h_{lm}$, and $Z_0$ are explicitly given as follows,
\begin{eqnarray}
    d_{lm}&=&\sqrt{\frac{(N-1+l+m)(N-2+l)!(N-2+m)!}{l!m!(N-1)!(N-2)!}},
    \label{dlm}
\\
    Z_0(\beta)&=&\int dv (v,w)H_{vwA}=\frac{(N-1)!}{(N\beta)^{N-1}}I_{N-1}(2N\beta),
\\
    h_{lm}(\beta)&=&\frac{1}{Z_0(\beta)d_{lm}}\int dv f_{lm}(v,w)H_{vwA}=\frac{I_{l+m+N-1}(2N\beta)}{I_{N-1}(2N\beta)},
\end{eqnarray}
where $I_n$ is the modified Bessel function of the first kind,
\begin{equation}
    I_n(x)=\sum_{q=0}^{\infty}\frac{1}{q!(q+n)!}\left(\frac{x}{2}\right)^{2q+n}.
    \label{mbes}
\end{equation}

\subsection{Character expansion of $\theta$ term}
Here let us see the character expansion of $\theta$ term in eq.\eqref{eq:Q}\cite{10.1143/ptp/93.1.161} 
and it can be transformed as
\begin{eqnarray}
    Q_{AA^\prime A^{\prime\prime}A^{\prime\prime\prime}}&=&\exp\left(\frac{\theta}{2\pi}\log \left[
    e^{i(A+A^\prime-A^{\prime\prime}-A^{\prime\prime\prime})}
    \right]\right)\\\nonumber
    &=&\int_{-\pi}^\pi dB\delta(B-[(A+A^\prime-A^{\prime\prime}-A^{\prime\prime\prime})\{\mathrm{mod}\ 2\pi\}])\exp(i\frac{\theta}{2\pi}B).
\end{eqnarray}
Next, we perform a Fourier series expansion,
\begin{equation}
    \exp(i\frac{\theta}{2\pi}B)=\sum_{p=-\infty}^\infty C_p(\theta)\exp(ipB),\quad C_p(\theta)=\frac{2(-1)^p\sin(\frac{\theta}{2})}{\theta-2\pi p},
    \label{C}
\end{equation}
and after integrating over $B$, we obtain
\begin{eqnarray}
    &&Q_{AA^\prime A^{\prime\prime}A^{\prime\prime\prime}}
    =\sum_{p=-\infty}^\infty C_p(\theta)\exp(ip[(A+A^\prime-A^{\prime\prime}-A^{\prime\prime\prime})]).
    \label{Qch}
\end{eqnarray}

\subsection{Construction of tensor network using character expansion}
\label{sec:TN_c}
In this subsection, we restrict our discussion to the CP(1) case, i.e., $N=2$.

\begin{figure}
    \centering
    \includegraphics[width=0.9\columnwidth]{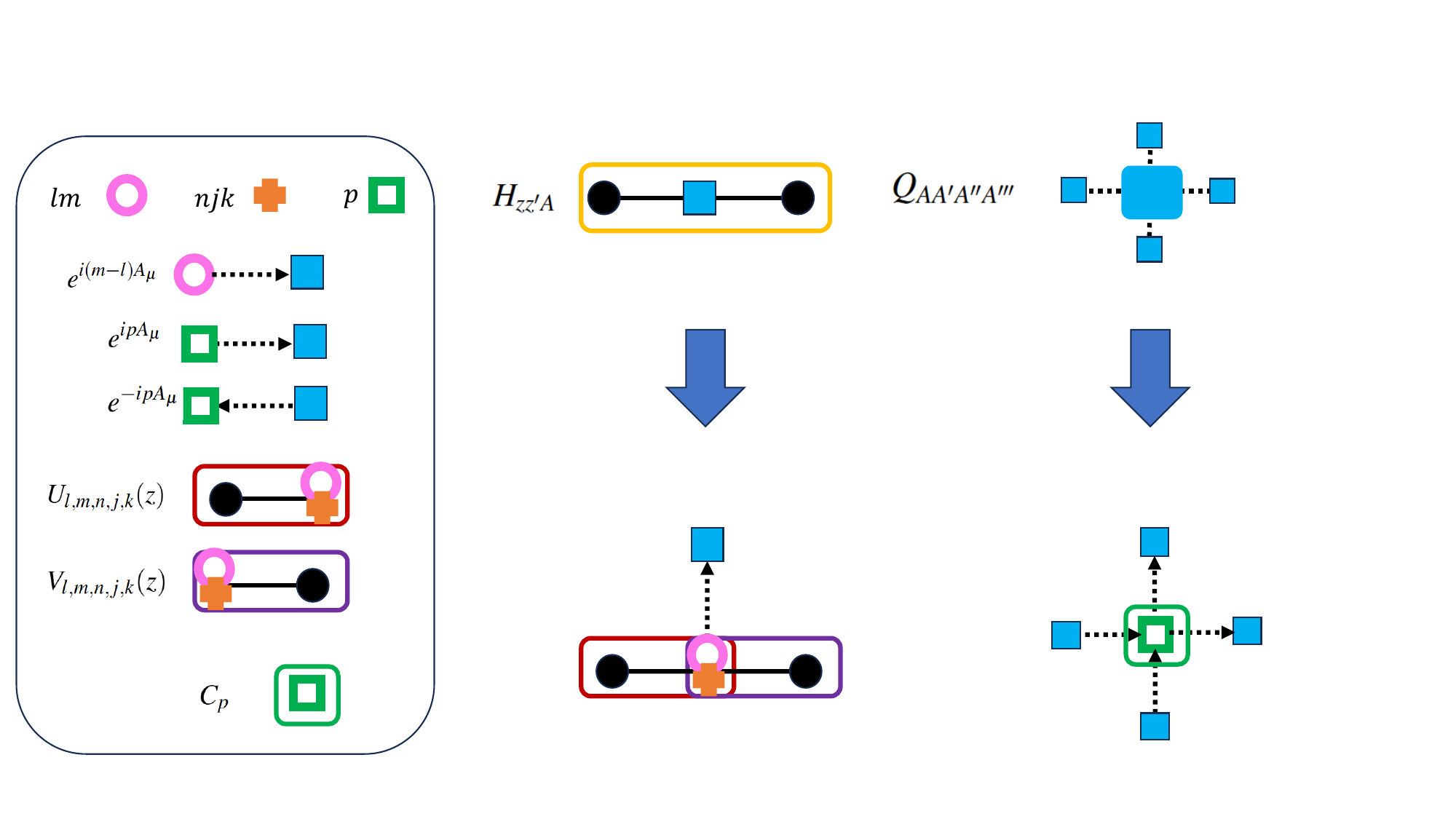}
    \caption{Graphical representation of the expansion of the $H$ and $Q$.
    Some building blocks of the expansion are summarized in the left box: indices are represented by open circle, open square and cross.
    Circle/square(s) with an arrow represent the phase factor say $e^{i(m-l)A_\mu}$.
    Circles/cross/square enclosed by a square box illustrate $U$, $V$ and $C_p$.
    The complex scalar field $z$ and the gauge field $A$ are represented as the black filled circle and blue filled square respectively as in Fig.~\ref{fig:tensor_quad}.
    The middle diagram shows the expansion of $H$ in Eq.~\eqref{chl} with the additional expansion for $f_{lm}$ in Eq.~\eqref{fUV}, while
    the right one is for the character expansion of $Q$ in Eq.~\eqref{Qch}.
    }
    \label{fig:chl_ch}
\end{figure}

In order to construct a tensor network representation, especially for the two-component complex scalar part, we need to separate\footnote{This separation differs from that used in previous studies \cite{Kawauchi:2017dnj,PhysRevD.105.054507}. For small $N$ (such as $N=2$ in the present case), our approach has an advantageous in constructing the initial tensor, because the dimensionality of the tensor increases more slowly compared to the previous method.}
 the orthogonal function $f_{lm}(v,w)$ in Eq.~\eqref{Flmndef}
  into functions of $v$ and $w$ respectively.
This can be achieved by applying the binomial theorem,
\begin{eqnarray}
    f_{lm}(v,w)&=&\sum_{n=0}^{\min(l,m)}F_{n}^{lm}(v\cdot w^*)^{l-n}(v^*\cdot w)^{m-n}\nonumber\\
    &=&\sum_{n=0}^{\min(l,m)}\sum_{j=0}^{l-n}\sum_{k=0}^{m-n}F_{n}^{lm}C^{l-n}_{j}C^{m-n}_{k}v_1^jv_1^{*k}v_2^{l-n-j}v_2^{*m-n-k}w_1^kw_1^{*j}w_2^{m-n-k}w_2^{*l-n-j},
\end{eqnarray}
with the combinatorial factor $C^n_m=\frac{n!}{(n-m)!m!}$.
Let $U$ and $V$ be defined as
\begin{eqnarray}
    U_{l,m,n,j,k}(v)&:=&F_{n}^{lm}C^{l-n}_{j}C^{m-n}_{k}v_1^jv_1^{*k}v_2^{l-n-j}v_2^{*m-n-k},
    \label{U}
\\
    V_{l,m,n,j,k}(w)&:=&w_1^kw_1^{*j}w_2^{m-n-k}w_2^{*l-n-j}.
    \label{V}
\end{eqnarray}
Then, $f_{lm}(v,w)$ can be written in terms of them
\begin{equation}
    f_{lm}(v,w)=\sum_{n,j,k} U_{l,m,n,j,k}(v) \, V_{l,m,n,j,k}(w).
\label{fUV}
\end{equation}
The graphical expression of the character-like expansion of $H$ in Eq.~\eqref{chl} together with the additional expansion for $f_{lm}$ in Eq.~\eqref{fUV} is shown in Fig.~\ref{fig:chl_ch}, where the character expansion of the $Q$ in Eq.~\eqref{Qch} is also illustrated.
\begin{figure}
    \centering
    \includegraphics[width=0.9\columnwidth]{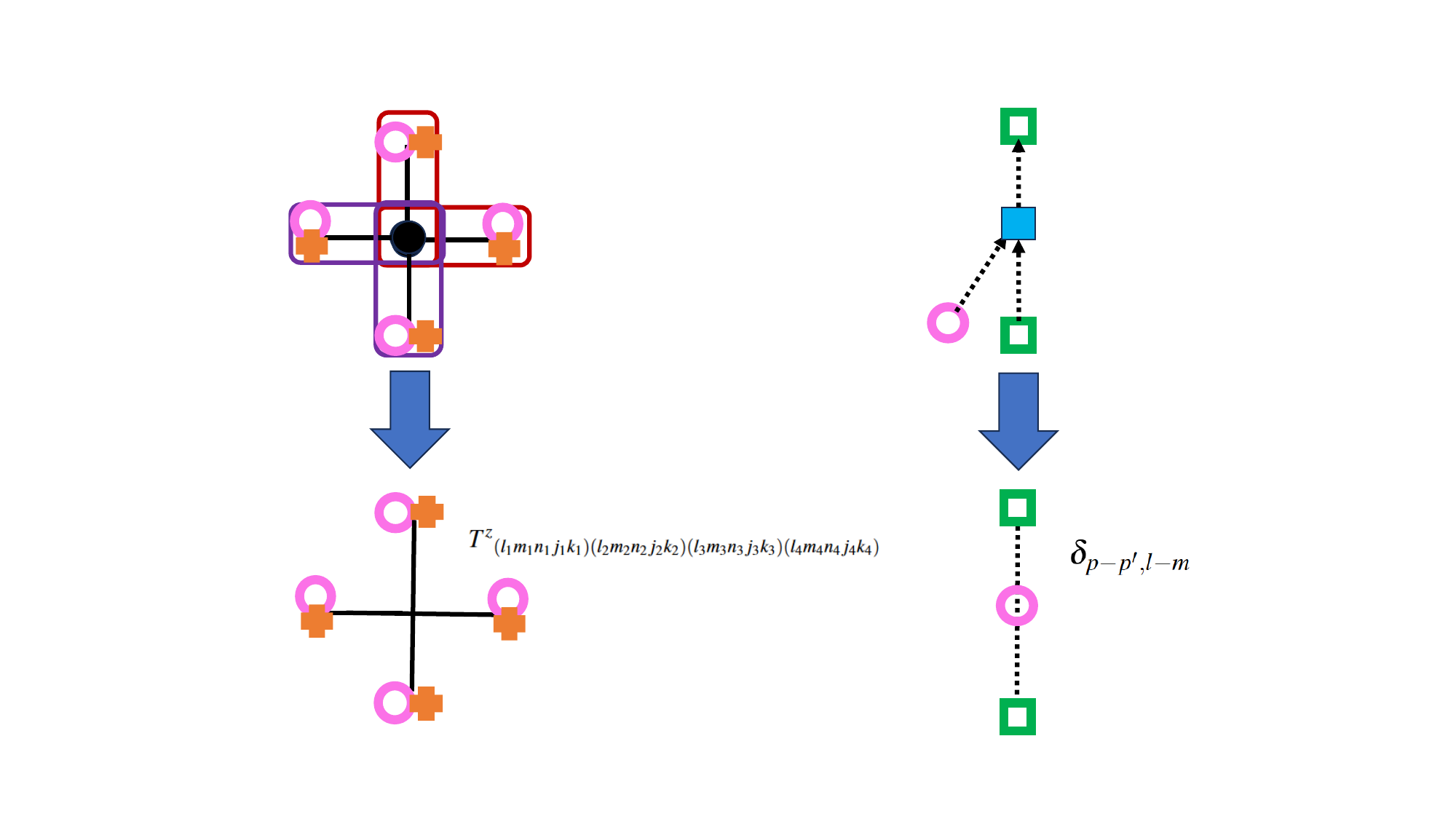}
    \caption{Left panel: Construction of $T^{\rm z}$ in eq.(\ref{eqn:Tz}) through the integration over $z$ (black filled circle).
    Note that after the integration the filled circle disappears.
    Right panel: Integration of $A_\mu$ (blue square) which yields a Kronecker delta.}
    \label{fig:tzdelta}
\end{figure}

After substituting the expansions of $H$ and $Q$ into the partition function in Eq.~\eqref{eq:ZHQ}, we then integrate over the scalar field $z$ and the gauge filed $A_\mu$, from which we shall obtain the tensor network representation.
First, let us see the integration over $z$.
For this purpose, the integral formula for the complex scalar field
\begin{equation}
	\int dz\,\,
    {z_1}^a{z_1}^{*b}{z_2}^c{z_2}^{*d}=\frac{a!c!}{(a+c+1)!}\delta_{a,b}\delta_{c,d},
    \label{intzform}
\end{equation}
is used to define a tensor $T^{\rm z}$ as follows,
\begin{eqnarray}
    &&{T^{\rm z}}_{(l_1m_1n_1j_1k_1)(l_2m_2n_2j_2k_2)(l_3m_3n_3j_3k_3)(l_4m_4n_4j_4k_4)}\nonumber\\
    &=&{Z_0}^2\sqrt{d_{l_1m_1}d_{l_2m_2}d_{l_3m_3}d_{l_4m_4}} \sqrt{h_{l_1m_1}h_{l_2m_2}h_{l_3m_3}h_{l_4m_4}}\nonumber\\
    &\times&\int dz U_{(l_1m_1n_1j_1k_1)}(z)U_{(l_2m_2n_2j_2k_2)}(z)V_{(l_3m_3n_3j_3k_3)}(z)V_{(l_4m_4n_4j_4k_4)}(z)\nonumber\\
    &=&{Z_0}^2\sqrt{d_{l_1m_1}d_{l_2m_2}d_{l_3m_3}d_{l_4m_4}} \sqrt{h_{l_1m_1}h_{l_2m_2}h_{l_3m_3}h_{l_4m_4}}\nonumber\\   
    &\times&F_{n_1}^{l_1m_1}C_{j_1}^{l_1-n_1}C_{k_1}^{m_1-n_1}F_{n_2}^{l_2m_2}C_{j_2}^{l_2-n_2}C_{k_2}^{m_2-n_2}
    \delta_{s,s^{\prime}}\delta_{u,u^{\prime}}\frac{s!(u-s-n)!}{(u-n+1)!},
\label{eqn:Tz}
\end{eqnarray}
where we have defined some new indices $s$, $s^\prime$, $u$, $u^\prime$, and $n$ as follows,
\begin{equation}
    s=j_1+j_2+k_3+k_4,
\end{equation}
\begin{equation}
    s^\prime=k_1+k_2+j_3+j_4,
\end{equation}
\begin{equation}
    u=l_1+l_2+m_3+m_4,
\end{equation}
\begin{equation}
    u^\prime=m_1+m_2+l_3+l_4,
\end{equation}
\begin{equation}
    n=n_1+n_2+n_3+n_4.
\end{equation}
The process of constructing $T^{\rm z}$ is illustrated in the left panel of Fig.~\ref{fig:tzdelta}.
In terms of $T^{\rm z}$, the partition function can be written as
\begin{eqnarray}
    Z&=&\prod_{x,\mu}(\sum_{l_{x,\mu}m_{x,\mu}n_{x,\mu}j_{x,\mu}k_{x,\mu}}\frac{1}{2\pi}\int_{-\pi}^{\pi} dA_\mu(x))\nonumber\\
    &\times&\prod_x\sum_{p_x}(C_{p_x}e^{ip_xA_0(x)}e^{-ip_{x-\hat{1}}A_0(x)}e^{-i(l_{x,0}-m_{x,0}) A_0(x)}e^{ip_{x-\hat{0}}A_1(x)}e^{-ip_xA_1(x)}e^{-i(l_{x,1}-m_{x,1})A_1(x)})\nonumber\\
    &\times&\prod_x {T^{\rm z}}_{(l_{x,0}m_{x,0}n_{x,0}j_{x,0}k_{x,0})(l_{x,1}m_{x,1}n_{x,1}j_{x,1}k_{x,1})
    (l_{x-\hat{0},0}m_{x-\hat{0},0}n_{x-\hat{0},0}j_{x-\hat{0},0}k_{x-\hat{0},0}) (l_{x-\hat{1},1}m_{x-\hat{1},1}n_{x-\hat{1},1}j_{x-\hat{1},1}k_{x-\hat{1},1})},
\end{eqnarray}

\begin{figure}
    \centering
    \includegraphics[width=0.9\columnwidth]{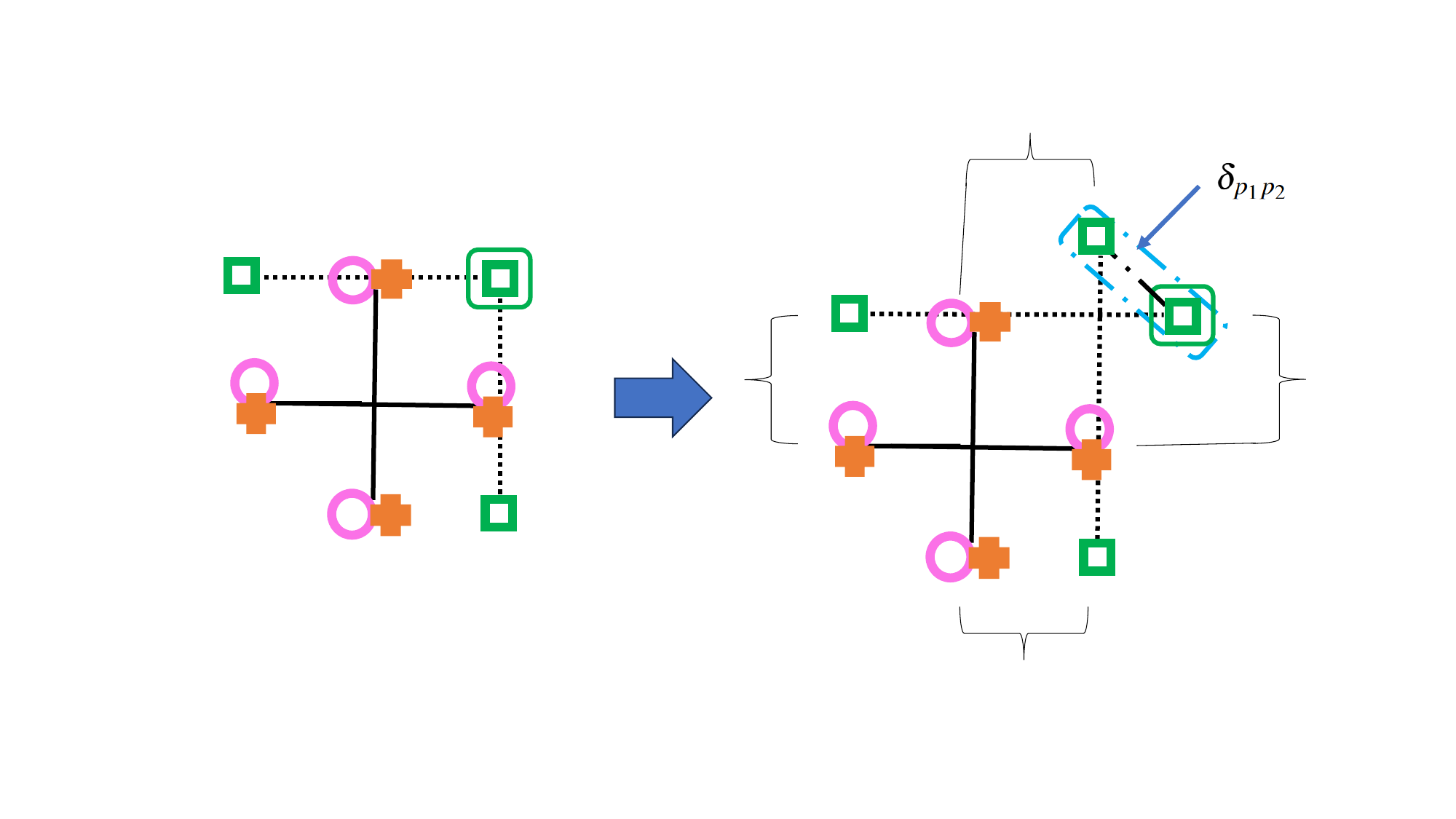}
    \caption{Constructing the initial tensor using the character expansions by inserting an additional Kronecker delta.}
    \label{fig:tensor_chl_ch}
\end{figure}

As the next step, we integrate over $A_\mu(x)$, as illustrated in the right panel of Fig.~\ref{fig:tzdelta}, and obtain
\begin{eqnarray}
    Z&=&\prod_{x,\mu}(\sum_{l_{x,\mu}m_{x,\mu}n_{x,\mu}j_{x,\mu}k_{x,\mu}})\prod_x\sum_{p_x}\nonumber\\
    &\times&\prod_x(C_{p_x}\delta_{p_x-p_{x-\hat{1}},l_{x,0}-m_{x,0}}\delta_{p_{x-\hat{0}}-p_x,l_{x,1}-m_{x,1}})\nonumber\\
    &\times&\prod_x {T^z}_{(l_{x,0}m_{x,0}n_{x,0}j_{x,0}k_{x,0})(l_{x,1}m_{x,1}n_{x,1}j_{x,1}k_{x,1})
    (l_{x-\hat{0},0}m_{x-\hat{0},0}n_{x-\hat{0},0}j_{x-\hat{0},0}k_{x-\hat{0},0})(l_{x-\hat{1},1}m_{x-\hat{1},1}n_{x-\hat{1},1}j_{x-\hat{1},1}k_{x-\hat{1},1})}.
\end{eqnarray}
In fact, the second line in the above equation is not suitable for constructing a conventional tensor network represention.
In order to construct the tensor network representation incorporating nearest-neighbor interactions,
we introduce an additional Kronecker delta $\delta_{p_1,p_2}$ as illustrated in Fig.~\ref{fig:tensor_chl_ch}. 
In the end, the partition function can be expressed in terms of the tensor network representation
\begin{eqnarray}
    Z&=&\prod_{x,\mu}(\sum_{l_{x,\mu}m_{x,\mu}n_{x,\mu}j_{x,\mu}k_{x,\mu}})\prod_x(\sum_{p_x}\sum_{p_x^\prime})\nonumber\\
    &\times&\prod_x{T}_{(l_{x,0}m_{x,0}n_{x,0}j_{x,0}k_{x,0}p_x)(l_{x,1}m_{x,1}n_{x,1}j_{x,1}k_{x,1}p_x^\prime)
    (l_{x-\hat{0},0}m_{x-\hat{0},0}n_{x-\hat{0},0}j_{x-\hat{0},0}k_{x-\hat{0},0}p_{x-\hat{0}})(l_{x-\hat{1},1}m_{x-\hat{1},1}n_{x-\hat{1},1}j_{x-\hat{1},1}k_{x-\hat{1},1}p_{x-\hat{1}})},\nonumber
\end{eqnarray}
where the initial tensor is given by
\begin{eqnarray}
    &&T_{(l_1m_1n_1j_1k_1p_1)(l_2m_2n_2j_2k_2p_2)(l_3m_3n_3j_3k_3p_3)(l_4m_4n_4j_4k_4p_4)}\nonumber\\
    &=&{Z_0}^2\sqrt{d_{l_1m_1}d_{l_2m_2}d_{l_3m_3}d_{l_4m_4}} \sqrt{h_{l_1m_1}h_{l_2m_2}h_{l_3m_3}h_{l_4m_4}}\nonumber\\   
    &\times&F_{n_1}^{l_1m_1}C_{j_1}^{l_1-n_1}C_{k_1}^{m_1-n_1}F_{n_2}^{l_2m_2}C_{j_2}^{l_2-n_2}C_{k_2}^{m_2-n_2}\nonumber\\
    &\times&\delta_{s,s^{\prime}}\delta_{u,u^{\prime}}\frac{s!(u-s-n)!}{(u-n+1)!}C_{p_1}\delta_{p_1,p_2}\delta_{p_2-p_4,l_1-m_1}\delta_{p_3-p_1,l_2-m_2}.
    \label{eqn:tensor_ch}
\end{eqnarray}
Note that all elements of the initial tensor are real-valued, even though the original action is complex number.
Of course, $C_p(\theta)$ and $F_n^{lm}$ can take negative value.

In a practical computation, one needs to truncate the expansions.
We denote the truncation order for the character-like expansion and the character expansion of $\theta$ term as $L_{\max}$ and $p_{\max}$, respectively.
Then, the total bond dimension of the initial tensor $\chi$ is now given by
a product of $\chi_{\rm z}$, the bond dimension of $T^{\rm z}$, and $\chi_{\rm A}$ for the $\theta$ term
\begin{equation}
\chi=\chi_{\rm z}\times\chi_{\rm A}
\end{equation}
with
\begin{eqnarray}
\chi_{\rm z}
&=&
\sum_{l,m=0}^{l+m \le L_{\max}}
\sum_{n=0}^{\min(l,m)}
\sum_{j=0}^{l-n}
\sum_{k=0}^{m-n}1,
\label{eqn:chi_z}
\\
\chi_{\rm A}
&=&
\sum_{p=-p_{\max}}^{p_{\max}-1}1
=
2p_{\max}.
\label{eqn:chi_A}
\end{eqnarray}

\section{Quadrature method on hypersphere}{\label{genz}}
 In this section, we describe the quadrature method on hyperspheres \cite{GENZ2003187}.
\subsection{Weights and sample points}
Let us define the integral of a function $f$ over the $(n-1)$-dimensional unit sphere $U_{n-1}=\{{\bm{x}}|{\bm{x}}\in\mathbb{R}^n, \sum_{i=1}^nx_i^2=1\}$ as
\begin{equation}
    I(f):=\int_{U_{n-1}}f({\bm x})d\sigma,
\end{equation}
where ${\bm{x}}=(x_1,x_2,\dots,x_n)\in U_{n-1}$, and $d\sigma$ denotes the normalized measure on $U_{n-1}$,
    $\int_{U_{n-1}}d\sigma=1$.
We approximate the integral by a quadrature formula,
\begin{equation}
    I(f)\approx
    I_m(f)
    :=
    \sum_{|{\bm{p}}|=m}
    w_{{\bm{p}}}
    f\{{\bm{u}}_{\bm{p}}\},
    \label{eqn:spherical_quadrature_formula_1}
\end{equation}
where the integer $m$ is the quadrature order, which serves as a control parameter: increasing its value is expected to improve the accuracy of the approximation.
Here, ${\bm p}$ is an $n$-component vector whose elements are non-negative integers
\begin{equation}
{\bm p}=(p_1,p_2,\dots,p_n),
\hspace{10mm}
p_i \in\{0,1,2,3,\dots,m\}
\end{equation}
and $|{\bm p}|$ denotes
\begin{equation}
|{\bm p}|
:=
p_1+p_2+\cdots+p_n.
\end{equation}
The notation $\sum_{|{\bm{p}}|=m}$ indicates the summation over all possible vectors $\bm{p}$ satisfying $|{\bm{p}}|=m$ for a given order $m$.
The total number of such configurations is given by the combinatorial factor $C^{m+n-1}_{m}$.

Now let us define the function $f\{{\bm x}\}$ in Eq.~\eqref{eqn:spherical_quadrature_formula_1} as
\begin{equation}
f\{{\bm x}\}
:=
\frac{1}{2^{c({\bm x})}}
\sum_{\{s\}}
f(s_1|x_1|,s_2|x_2|,\dots,s_n|x_n|),
\label{eqn:f2}
\end{equation}
where $c(\bm{x})$ is the number of nonzero components of $\bm{x}$,
and the summation $\sum_{\{s\}}$ runs over all combinations of $s_i=\pm1$ for all $i$ with $x_i\neq0$.
In other words, this definition symmetrizes the function with respect to individual sign flips of the nonzero components of $\bm{x}$.
Since the integral of the odd part of a function over a hypersphere vanishes, it suffices to consider only the even part.

The weight $w_{{\bm p}}$ is given by
\begin{equation}
w_{{\bm p}}
=
I\left(
\prod_{i=1}^n
\prod_{j=0}^{p_i-1}
\frac{x_i^2-u_j^2}{u_{p_i}^2-u_j^2}
\right)
=
\int_{U_{n-1}}
\left[
\prod_{i=1}^n
\prod_{j=0}^{p_i-1}
\frac{x_i^2-u_j^2}{u_{p_i}^2-u_j^2}
\right]
d\sigma,
\label{eqn:weight}
\end{equation}
where $u_i$ are defined, for $i=0,1,2,\dots,n$, by
\begin{equation}
u_i:=
\sqrt{\frac{i+\mu}{m+\mu n}}.
\label{eqn:u_i}
\end{equation}
Here, $\mu$ is a free parameter in the range $0 \le \mu \le 1$. 
Setting $\mu$ to zero reduces the number of integration points, 
but it is generally recommended to choose a small finite value to avoid numerical instability.
Using the $u_i$ in Eq.~(\ref{eqn:u_i}), the vector ${\bm u}_{\bm p}$ appearing in the argument of $f$ in Eq.~\eqref{eqn:spherical_quadrature_formula_1} is defined as the $n$-component vector
\begin{equation}
{\bm u}_{\bm p}
=
(u_{p_1},u_{p_2},...,u_{p_n}).
\end{equation}
To explicitly compute the weights in Eq.~\eqref{eqn:weight}, the following formula is useful,
\begin{equation}
\Phi(k_1,k_2,\dots,k_n)
:=
\int_{U_{n-1}} x_1^{2k_1}x_2^{2k_2}\cdots x_n^{2k_n}d\sigma=\frac{\Gamma(\frac{n}{2})}{\pi^{\frac{n}{2}}}\frac{\prod_{i=1}^n \Gamma(k_i+\frac{1}{2})}{\Gamma(|\bm{k}|+\frac{n}{2})},
\label{eq:intsn}
\end{equation}
where $\Gamma(x)$ is the gamma function.
An example for $n=3$ is provided in the next subsection.

Equations \eqref{eqn:spherical_quadrature_formula_1} with \eqref{eqn:f2} can be rewritten in a compact form
\begin{equation}
    \int_{U_2}d\sigma f(\bm{x})\approx\sum_{i=1}^{N_{\rm{z}}} w_i f(\bm{x}_i),
\end{equation}
where the index $i$ runs over all combinations of $(\bm{p},\bm{s})=i$, and
\begin{eqnarray}
	N_{\rm z}&=&C^{m+n-1}_{m}\times2^{n} \hspace{5mm}(\mbox{for }\mu\neq0),
\label{eqn:Nz_general}
	\\
    w_i&=&\frac{w_{\bm{p}}}{2^{c(\bm{u}_{\bm{p}})}}=\frac{w_{\bm{p}}}{2^{n}} \hspace{5mm}(\mbox{for }\mu\neq0),
\label{eqn:Wz}
\\
    f(\bm{x}_i)&=&f_{\bm{p},\bm{s}}=f(s_1|u_{p_1}|,s_2|u_{p_2}|,\dots,s_n|u_{p_n}|).
\label{eqn:sample_fx}
\end{eqnarray}

\subsection{Example for $n=3$}
Examples for $n=3$ with $m=1,2$ are shown below.
The weights $w_{\bm{p}}$ can be computed using Eq.~\eqref{eq:intsn}.
\begin{itemize}
\item $m=1=|{\bm p}|$, $C_1^3=3$
\begin{equation}
    \bm{p}=(p_1,p_2,p_3)
    =
    (1,0,0),(0,1,0),(0,0,1).
\end{equation}
\begin{eqnarray}
    u_i&=&\sqrt{\frac{i+\mu}{1+3\mu}}\,\,\, \mbox{ for} \,\,\, i=0,1.
\end{eqnarray}
\begin{eqnarray}
    \bm{u}_{(1,0,0)}&=&\left(\sqrt{\frac{1+\mu}{1+3\mu}},\sqrt{\frac{\mu}{1+3\mu}},\sqrt{\frac{\mu}{1+3\mu}}\right),\nonumber\\
    \bm{u}_{(0,1,0)}&=&\left(\sqrt{\frac{\mu}{1+3\mu}},\sqrt{\frac{1+\mu}{1+3\mu}},\sqrt{\frac{\mu}{1+3\mu}}\right),\nonumber\\
    \bm{u}_{(0,0,1)}&=&\left(\sqrt{\frac{\mu}{1+3\mu}},\sqrt{\frac{\mu}{1+3\mu}},\sqrt{\frac{1+\mu}{1+3\mu}}\right).
\end{eqnarray}
Since all components of $\bm{u}_{\bm{p}}$ are nonzero when $\mu\neq0$, in total there are $3\times2^3=24$ sample points.
\begin{eqnarray}
    w_{(1,0,0)}
    &=&
    \int_{U_2}
    \prod_{j=0}^{p_1-1}
    \frac{x_1^2-u_j^2}{u_{p_1}^2-u_j^2}
    d\sigma
    =
    \int_{U_2}
    \frac{x_1^2-u_0^2}{u_1^2-u_0^2}
    d\sigma
    =
    \int_{U_2}
    \frac{x_1^2-\frac{\mu}{1+3\mu}}{\frac{1}{1+3\mu}}
    d\sigma\nonumber\\
    &=&
    \int_{U_2}
    \left[
    (1+3\mu)x_1^2-\mu
    \right]
    d\sigma
    =(1+3\mu)\Phi(1,0,0)-\mu,
    \nonumber\\
    w_{(0,1,0)}
    &=&
    =(1+3\mu)\Phi(0,1,0)-\mu,
    \nonumber\\
    w_{(0,0.1)}
    &=&
    =(1+3\mu)\Phi(0,0,1)-\mu.
\end{eqnarray}

\item $m=2=|{\bm p}|$, $C^4_2=6$
\begin{eqnarray}
    \bm{p}=(p_1,p_2,p_3)
    &=&
    (2,0,0),(0,2,0),(0,0,2),
    (1,1,0),(1,0,1),(0,1,1).
\end{eqnarray}
\begin{eqnarray}
    u_i&=&\sqrt{\frac{i+\mu}{2+3\mu}}\,\,\, \mbox{ for} \,\,\, i=0,1,2.
\end{eqnarray}
\begin{eqnarray}
    \bm{u}_{(2,0,0)}&=&\left(\sqrt{\frac{2+\mu}{2+3\mu}},\sqrt{\frac{\mu}{2+3\mu}},\sqrt{\frac{\mu}{2+3\mu}}\right),\nonumber\\
    \bm{u}_{(0,2,0)}&=&\left(\sqrt{\frac{\mu}{2+3\mu}},\sqrt{\frac{2+\mu}{2+3\mu}},\sqrt{\frac{\mu}{2+3\mu}}\right),\nonumber\\
    \bm{u}_{(0,0,2)}&=&\left(\sqrt{\frac{\mu}{2+3\mu}},\sqrt{\frac{\mu}{2+3\mu}},\sqrt{\frac{2+\mu}{2+3\mu}}\right),\nonumber\\
    \bm{u}_{(1,1,0)}&=&\left(\sqrt{\frac{1+\mu}{2+3\mu}},\sqrt{\frac{1+\mu}{2+3\mu}},\sqrt{\frac{\mu}{2+3\mu}}\right),\nonumber\\
    \bm{u}_{(1,0,1)}&=&\left(\sqrt{\frac{1+\mu}{2+3\mu}},\sqrt{\frac{\mu}{2+3\mu}},\sqrt{\frac{1+\mu}{2+3\mu}}\right),\nonumber\\
    \bm{u}_{(0,1,1)}&=&\left(\sqrt{\frac{\mu}{2+3\mu}},\sqrt{\frac{1+\mu}{2+3\mu}},\sqrt{\frac{1+\mu}{2+3\mu}}\right).
\end{eqnarray}
The number of the sample points for $\mu\neq0$ is $6\times2^3=48$.
\begin{eqnarray}
    w_{(2,0,0)}
    &=&
    \int_{U_2}
    \left[
    \prod_{j=0}^{p_1-1}
    \frac{x_1^2-u_j^2}{u_{p_1}^2-u_j^2}
    \right]
    d\sigma
    =
    \int_{U_2}
    \left[
    \frac{(x_1^2-u_0^2)(x_1^2-u_1^2)}{(u_2^2-u_0^2)(u_2^2-u_1^2)}
    \right]
    d\sigma
    \nonumber\\
    &=&
    \int_{U_2}
    \left[
    \frac{(x_1^2-\frac{\mu}{2+3\mu})(x_1^2-\frac{1+\mu}{2+3\mu})}{\frac{2}{(2+3\mu)^2}}
    \right]
    d\sigma
    \nonumber\\
    &=&
    \frac{1}{2}
    \left[
    (2+3\mu)^2\Phi(2,0,0)
    -(1+2\mu)(2+3\mu)\Phi(1,0,0)
    +\mu+\mu^2\right],
    \nonumber\\
    w_{(0,2,0)}
    &=&
    \frac{1}{2}
    \left[
    (2+3\mu)^2\Phi(0,2,0)
    -(1+2\mu)(2+3\mu)\Phi(0,1,0)
    +\mu+\mu^2\right],
    \nonumber\\
    w_{(0,0.2)}
    &=&
    \frac{1}{2}
    \left[
    (2+3\mu)^2\Phi(0,0,2)
    -(1+2\mu)(2+3\mu)\Phi(0,0,1)
    +\mu+\mu^2\right],
    \nonumber\\
    w_{(1,1,0)}
    &=&
    \int_{U_2}
    \left[
    \prod_{j=0}^{p_1-1}
    \frac{x_1^2-u_j^2}{u_{p_1}^2-u_j^2}
    \prod_{j=0}^{p_2-1}
    \frac{x_2^2-u_j^2}{u_{p_2}^2-u_j^2}
    \right]
    d\sigma
    =
    \int_{U_2}
    \left[
    \frac{(x_1^2-u_0^2)(x_2^2-u_0^2)}{(u_1^2-u_0^2)(u_1^2-u_0^2)}
    \right]d\sigma
    \nonumber\\
    &=&
    (2+3\mu)^2\Phi(1,1,0)-\mu(2+3\mu)(\Phi(1,0,0)+\Phi(0,1,0))+\mu^2,
    \nonumber\\
    w_{(1,0,1)}
    &=&
    (2+3\mu)^2\Phi(1,0,1)-\mu(2+3\mu)(\Phi(1,0,0)+\Phi(0,0,1))+\mu^2,
    \nonumber\\
    w_{(0,1,1)}
    &=&
    (2+3\mu)^2\Phi(0,1,1)-\mu(2+3\mu)(\Phi(0,1,0)+\Phi(0,0,1))+\mu^2.
\end{eqnarray}
\end{itemize}

\section{Exact results for $2\times2$ lattice}\label{ex}
\label{sec:exact_2x2}

For sufficiently small lattice, such as $2\times2$ sites, the corresponding partition function can be numerically evaluated by brute force with an accuracy of $O(10^{-10})$, as we explain below.
Here we discuss the general CP($N-1$) model, while numerical results are presented for $N=2$.

\begin{figure}
    \centering
    \includegraphics[width=0.3\columnwidth]{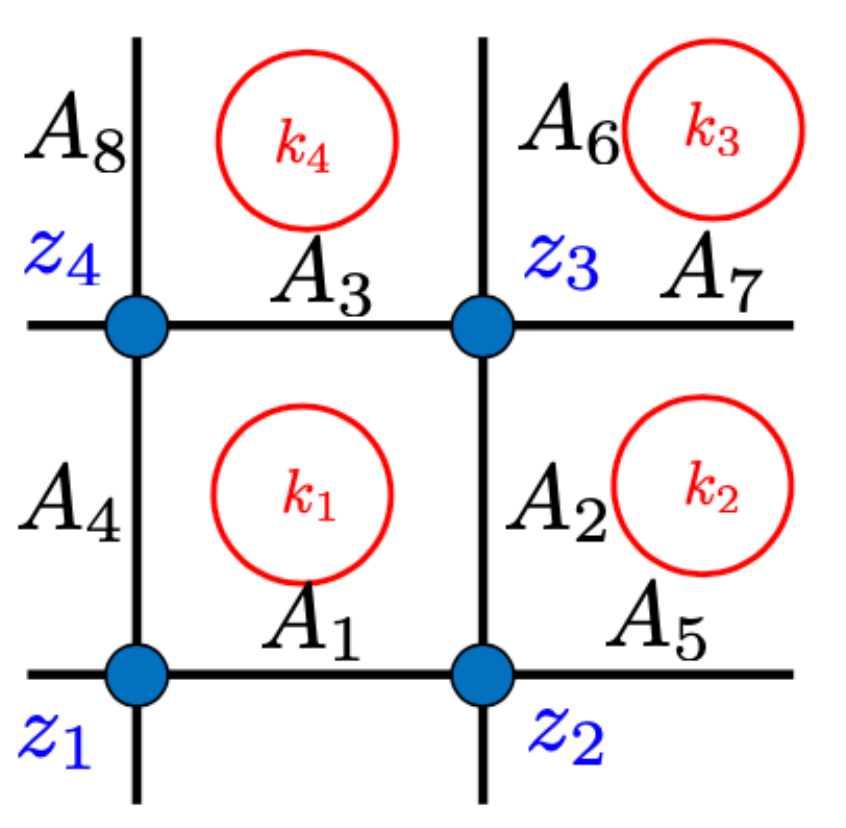}
    \caption{Assignment of the scalar fields and the gauge fields for $Z_{2\times2}$ in Eq~\eqref{eqn:Z_beta_theta}.}
    \label{fig:Z_2x2}
\end{figure}

From Eq.~\eqref{eq:ZHQ}, the partition function for the $2\times2$ lattice with PBC is given explicitly as
\begin{eqnarray}
    Z_{2\times2}(\beta,\theta)
    &=&
    \int dz_1dz_2dz_3dz_4
    \left(\frac{1}{2\pi}\right)^8\int dA_1dA_2dA_3dA_4dA_5dA_6dA_7dA_8
    \nonumber\\
    &&
    H_{z_1z_2A_1}H_{z_2z_1A_5}
    H_{z_2z_3A_2}H_{z_3z_2A_6}
    H_{z_4z_3A_3}H_{z_3z_4A_7}
    H_{z_1z_4A_4}H_{z_4z_1A_8}
    \nonumber\\
    &\times&
    Q_{A_1A_2A_3A_4}
    Q_{A_5A_4A_7A_2}
    Q_{A_7A_8A_5A_6}
    Q_{A_3A_6A_1A_8}.
    \label{eqn:Z_beta_theta}
\end{eqnarray}
See Fig.~\ref{fig:Z_2x2} for the field assignment.
Using Eq.~\eqref{Qch} and integrating over the gauge field, it is written as follows,
\begin{eqnarray}
    \lefteqn{Z_{2\times2}(\beta,\theta)}\\
    &=&
    \sum_{k_1\in\mathbb{Z}}
    \sum_{k_2\in\mathbb{Z}}
    \sum_{k_3\in\mathbb{Z}}
    \sum_{k_4\in\mathbb{Z}}
    C_{k_1}(\theta)
    C_{k_2}(\theta)
    C_{k_3}(\theta)
    C_{k_4}(\theta)
    \int dz_1dz_2dz_3dz_4
    \nonumber\\
    &&
    G_{k_1-k_4,k_2-k_3}(z_1,z_2)
    G_{k_1-k_2,k_4-k_3}(z_2,z_3)
    G_{k_1-k_4,k_2-k_3}(z_3,z_4)
    G_{k_1-k_2,k_4-k_3}(z_4,z_1),
\label{eqn:Z_ch2x2}
\end{eqnarray}
where $G_{k,k\prime}$ is defined as
\begin{equation}    
    G_{k,k^\prime}(z,z^\prime):= F_k(z,z^\prime)F_{k^\prime}(z^\prime,z),
\end{equation}
with $F_k$ defined as
\begin{equation}
    F_k(z,z^\prime):=\int_{-\pi}^\pi \frac{dA}{2\pi}H_{zz^\prime A}e^{ikA}=e^{-ik\cdot{\rm arg}[z^\dag z^\prime]}I_k(2\beta N|z^\dag z^\prime|).
    \label{integrateA}
\end{equation}
Here, $I_k(x)$ is the modified Bessel function of the first kind.

The integrals of $z$ can be done by using the quadrature method in Eq.~\eqref{eq:Nz}, and then it turns out to be matrix product operation,
\begin{eqnarray}
\lefteqn{
    \int dz_1dz_2dz_3dz_4
    G_{k_1-k_4,k_2-k_3}(z_1,z_2)
    G_{k_1-k_2,k_4-k_3}(z_2,z_3)
    G_{k_1-k_4,k_2-k_3}(z_3,z_4)
    G_{k_1-k_2,k_4-k_3}(z_4,z_1)}\nonumber\\
    &\approx&
    \sum_{z_1,z_2,z_3,z_4}W^{({\rm z})}_{z_1}W^{({\rm z})}_{z_2}W^{({\rm z})}_{z_3}W^{({\rm z})}_{z_4}\nonumber\\
    &\times&
    G_{k_1-k_4,k_2-k_3}(z_1,z_2)
    G_{k_1-k_2,k_4-k_3}(z_2,z_3)
    G_{k_1-k_4,k_2-k_3}(z_3,z_4)
    G_{k_1-k_2,k_4-k_3}(z_4,z_1).
\end{eqnarray}
The summation in the character expansions of Eq.~\eqref{eqn:Z_ch2x2} can be performed efficiently using the following variable transformation,
\begin{eqnarray}
    k_1&=&n+l_1,
    \\
    k_2&=&n+l_2,
    \\
    k_3&=&n+l_3,
    \\
    k_4&=&n.
\end{eqnarray}
Then, the summations are transformed as
\begin{equation}
    \sum_{k_1\in\mathbb{Z}}
    \sum_{k_2\in\mathbb{Z}}
    \sum_{k_3\in\mathbb{Z}}
    \sum_{k_4\in\mathbb{Z}}
    \ldots
    =
    \sum_{n\in\mathbb{Z}}\,\,
    \sum_{(l_1,l_2,l_3)\in\mathbb{Z}^3}\ldots.
\end{equation}
After the transformation, the partition function is written as
\begin{equation}
    Z_{2\times2}(\beta,\theta)
    =
    \sum_{(l_1,l_2,l_3)}
    c_{l_1l_2l_3}(\theta)
    g_{l_1l_2l_3}(\beta),
\label{eqn:Z_2x2_final}
\end{equation}
with
\begin{eqnarray}
    c_{l_1l_2l_3}(\theta)
    &=&
    \sum_{n\in\mathbb{Z}}
    C_{n+l_1}(\theta)
    C_{n+l_2}(\theta)
    C_{n+l_3}(\theta)
    C_{n}(\theta),
\label{eqn:cC}
    \\
    g_{l_1l_2l_3}(\beta)
    &=&
    \sum_{z_1,z_2,z_3,z_4}W^{({\rm z})}_{z_1}W^{({\rm z})}_{z_2}W^{({\rm z})}_{z_3}W^{({\rm z})}_{z_4}
    G_{l_1,l_2 -l_3}
    G_{l_1-l_2,-l_3}
    G_{l_1,l_2 -l_3}
    G_{l_1-l_2,-l_3}.
\label{eqn:gG}
\end{eqnarray}
In practical calculations, one has to truncate the summations in Eq.~\eqref{eqn:Z_2x2_final}-\eqref{eqn:cC}, and the integration order $m$ (or $N_{\rm z}$) in Eq.~\eqref{eqn:gG} should be taken sufficiently large.
We carefully examine the choice of the truncation order and estimate the associated errors.
The numerical results of $f_{\rm 2\times2}=-\frac{1}{4}\ln Z_{2\times2}$ for $N=2$ with several sets of $(\beta,\theta)$ are presented in Table~\ref{tab:Z_2x2}, together with the estimated errors.
\begin{table}[h]
\centering
\caption{Exact results of $f_{2\times2}(\beta,\theta)$ with $N=2$}
\begin{tabular}{l|l|l|l|l}
\diagbox{$\beta$}{$\theta$} & $0$ & $0.5\pi$ & $0.9\pi$ & $\pi$ \\\hline
0.1 & $-$0.039998271459(1) & \phantom{$-$}0.06127846045(1) & \phantom{$-$}0.22237598844(1) & \phantom{$-$}0.2342968125(1) \\
0.5 & $-$0.98397311020(1)  & $-$0.91398453764(1) & $-$0.82221041724(1) & $-$0.81643793149(1) \\
1.0 & $-$3.5442323086(1)   & $-$3.5346034415(1)  & $-$3.52508840794(1) & $-$3.5245888065(1) \\
\end{tabular}
\label{tab:Z_2x2}
\end{table}

\section{Numerical investigation of SO(3) and CP symmetry breaking}
\label{sec:symmetry}
We assess the extent to which the SO(3) symmetry and the CP symmetry at $\theta=\pi$ are violated due to the quadrature method and the coarse-graining procedure.
To this end we compute the corresponding order parameters for each symmetry.
In the following numerical results, the parameters for the initial tensor, and the coarse-graining algorithm are identical to those used in the main text.

\begin{figure}[h]
    \centering
    \includegraphics[width=0.7\columnwidth]{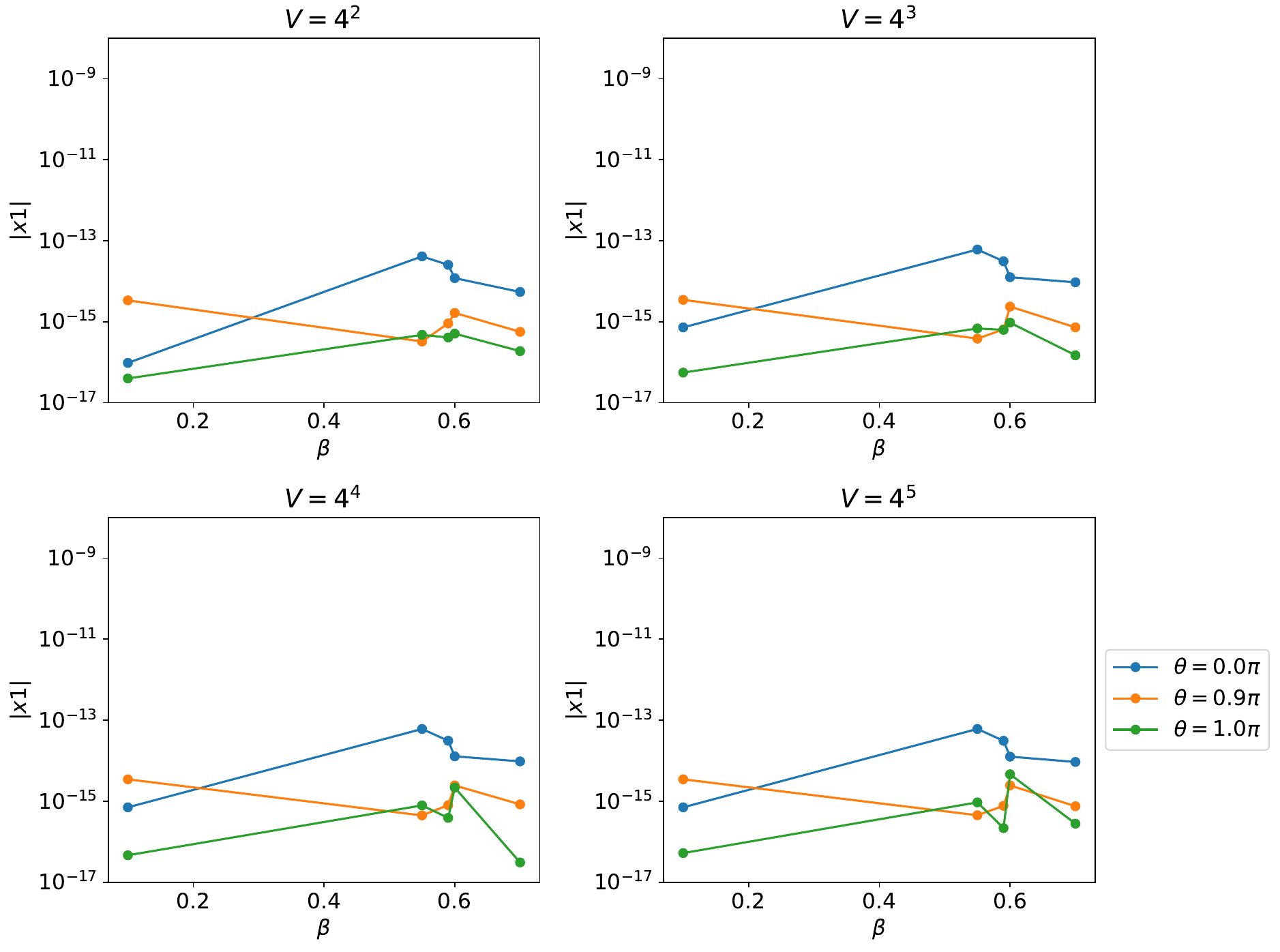}
    \caption{The order parameter for the SO(3) symmetry, $x_1$, as a function of $\beta$ for $V=4^2-4^5$. 
    }
    \label{x1}
\end{figure}

The order parameter associated with the SO(3) symmetry may be defined as
\begin{equation}
   x_i:= \braket{z^{\dagger}\sigma_iz} ,
\end{equation}
where $z$ is the complex scalar field and $\sigma_i$ ($i=1,2,3$) denote the Pauli matrices.
To evaluate this observable, we employ the impurity tensor network technique\cite{Morita:2024lwg}.
Figure~\ref{x1} shows the numerical results for $x_1 = \braket{z^{\dagger}\sigma_1z}$.
For all values of $\beta$, $\theta$, and the volume $V$ shown in the figure, the SO(3) breaking remains below $O(10^{-12})$ and exhibits negligible volume dependence. 
We therefore conclude that the violation of the SO(3) symmetry is well controlled, and
no strongly relevant perturbation is induced by the numerical procedures.

\begin{figure}[h]
    \centering
    \includegraphics[width=0.7\columnwidth]{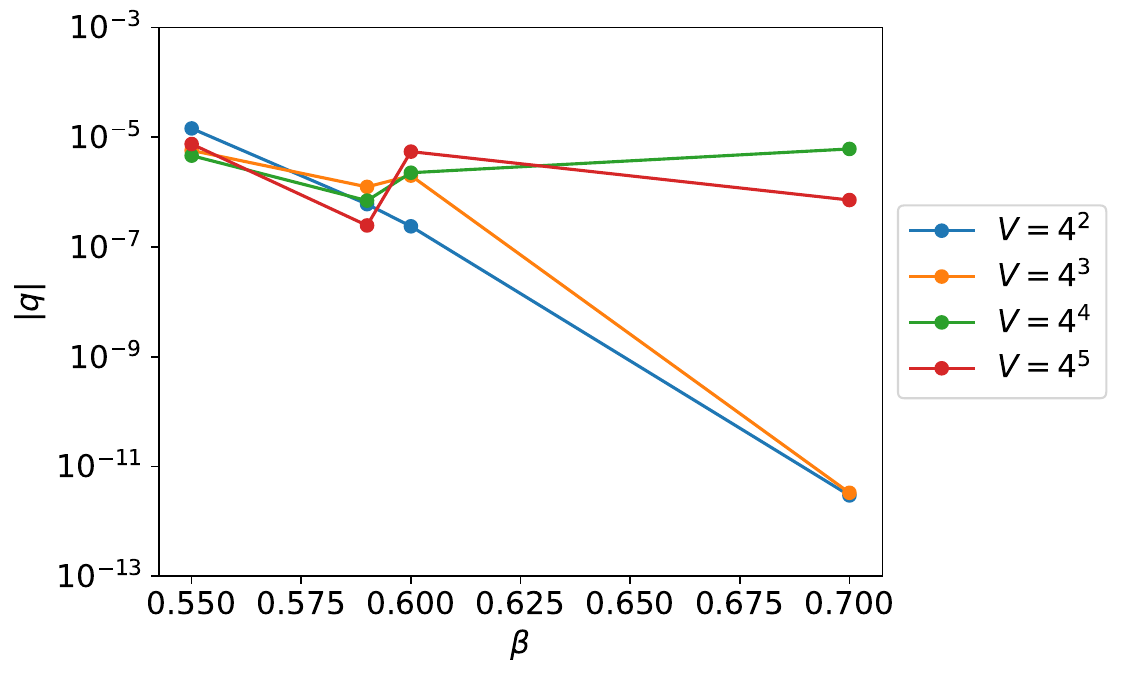}
    \caption{The order parameter for the CP symmetry, the topological charge density $q$, as a function of $\beta$ for $V=4^2-4^5$ at $\theta=\pi$.}
    \label{cpimp}
\end{figure}

As an order parameter for CP symmetry, we consider the topological charge density
\begin{equation}
    q:= \frac{\partial f}{\partial\theta}, 
\end{equation}
where $f$ is the free energy given in Eq.~\eqref{eqn:f}.
We evaluate $q$ using the impurity method as well.
Figure~\ref{cpimp} presents the numerical results for $q$ at $\theta=\pi$ in the region $\beta_{\rm c}\lesssim\beta$,
where CP symmetry implies that $q$ should vanish.
Our results show that $q$ remains smaller than $O(10^{-5})$. 
Although this breaking appears larger than that observed for the SO(3) symmetry, the volume dependence around $\beta_{\rm c}\approx0.6$ is weak.
Thus the CP symmetry is expected to be under sufficient control within our numerical setup.

\clearpage
\bibliography{CP(1)}

\providecommand{\href}[2]{#2}\begingroup\raggedright\begin{thebibliography}{10}

\bibitem{Callan:1976je}
C.G.~Callan, Jr., R.F.~Dashen and D.J.~Gross, \emph{{The Structure of the Gauge
  Theory Vacuum}},
  \href{https://doi.org/10.1016/0370-2693(76)90277-X}{\emph{Phys. Lett. B}
  {\bfseries 63} (1976) 334}.

\bibitem{PhysRevLett.94.170201}
M.~Troyer and U.-J.~Wiese, \emph{Computational complexity and fundamental
  limitations to fermionic quantum monte carlo simulations},
  \href{https://doi.org/10.1103/PhysRevLett.94.170201}{\emph{Phys. Rev. Lett.}
  {\bfseries 94} (2005) 170201}.

\bibitem{CREWTHER1979123}
R.~Crewther, P.~{Di Vecchia}, G.~Veneziano and E.~Witten, \emph{Chiral estimate
  of the electric dipole moment of the neutron in quantum chromodynamics},
  \href{https://doi.org/https://doi.org/10.1016/0370-2693(79)90128-X}{\emph{Physics
  Letters B} {\bfseries 88} (1979) 123}.

\bibitem{PhysRevLett.124.081803}
C.~Abel, S.~Afach, N.J.~Ayres, C.A.~Baker, G.~Ban, G.~Bison et~al.,
  \emph{Measurement of the permanent electric dipole moment of the neutron},
  \href{https://doi.org/10.1103/PhysRevLett.124.081803}{\emph{Phys. Rev. Lett.}
  {\bfseries 124} (2020) 081803}.

\bibitem{PhysRevD.90.074503}
Y.~Shimizu and Y.~Kuramashi, \emph{Critical behavior of the lattice schwinger
  model with a topological term at $\ensuremath{\theta}=\ensuremath{\pi}$ using
  the grassmann tensor renormalization group},
  \href{https://doi.org/10.1103/PhysRevD.90.074503}{\emph{Phys. Rev. D}
  {\bfseries 90} (2014) 074503}.

\bibitem{10.1093/ptep/ptv022}
S.~Takeda and Y.~Yoshimura, \emph{Grassmann tensor renormalization group for
  the one-flavor lattice gross-neveu model with finite chemical potential},
  \href{https://doi.org/10.1093/ptep/ptv022}{\emph{Progress of Theoretical and
  Experimental Physics} {\bfseries 2015} (2015) 043B01}
  [\href{https://arxiv.org/abs/https://academic.oup.com/ptep/article-pdf/2015/4/043B01/19301062/ptv022.pdf}{{\ttfamily
  https://academic.oup.com/ptep/article-pdf/2015/4/043B01/19301062/ptv022.pdf}}].

\bibitem{Kawauchi:2017dnj}
H.~Kawauchi and S.~Takeda, \emph{{Loop-TNR analysis of CP(1) model with theta
  term}}, \href{https://doi.org/10.1051/epjconf/201817511015}{\emph{EPJ Web
  Conf.} {\bfseries 175} (2018) 11015}
  [\href{https://arxiv.org/abs/1710.09804}{{\ttfamily 1710.09804}}].

\bibitem{Kadoh:2018hqq}
D.~Kadoh, Y.~Kuramashi, Y.~Nakamura, R.~Sakai, S.~Takeda and Y.~Yoshimura,
  \emph{{Tensor network formulation for two-dimensional lattice $ \mathcal{N} $
  = 1 Wess-Zumino model}},
  \href{https://doi.org/10.1007/JHEP03(2018)141}{\emph{JHEP} {\bfseries 03}
  (2018) 141} [\href{https://arxiv.org/abs/1801.04183}{{\ttfamily
  1801.04183}}].

\bibitem{PhysRevD.105.054507}
K.~Nakayama, L.~Funcke, K.~Jansen, Y.-J.~Kao and S.~K\"uhn, \emph{Phase
  structure of the {{CP(1)}} model in the presence of a topological
  $\ensuremath{\theta}$-term},
  \href{https://doi.org/10.1103/PhysRevD.105.054507}{\emph{Phys. Rev. D}
  {\bfseries 105} (2022) 054507}
  [\href{https://arxiv.org/abs/2107.14220}{{\ttfamily 2107.14220}}].

\bibitem{Akiyama:2020soe}
S.~Akiyama, Y.~Kuramashi, T.~Yamashita and Y.~Yoshimura, \emph{{Restoration of
  chiral symmetry in cold and dense Nambu--Jona-Lasinio model with tensor
  renormalization group}},
  \href{https://doi.org/10.1007/JHEP01(2021)121}{\emph{JHEP} {\bfseries 01}
  (2021) 121} [\href{https://arxiv.org/abs/2009.11583}{{\ttfamily
  2009.11583}}].

\bibitem{10.1093/ptep/ptac014}
S.~Akiyama, Y.~Kuramashi and T.~Yamashita, \emph{Metal-insulator transition in
  the (2+1)-dimensional hubbard model with the tensor renormalization group},
  \href{https://doi.org/10.1093/ptep/ptac014}{\emph{Progress of Theoretical and
  Experimental Physics} {\bfseries 2022} (2022) 023I01}
  [\href{https://arxiv.org/abs/https://academic.oup.com/ptep/article-pdf/2022/2/023I01/42931059/ptac014.pdf}{{\ttfamily
  https://academic.oup.com/ptep/article-pdf/2022/2/023I01/42931059/ptac014.pdf}}].

\bibitem{PhysRevLett.99.120601}
M.~Levin and C.P.~Nave, \emph{Tensor renormalization group approach to
  two-dimensional classical lattice models},
  \href{https://doi.org/10.1103/PhysRevLett.99.120601}{\emph{Phys. Rev. Lett.}
  {\bfseries 99} (2007) 120601}
  [\href{https://arxiv.org/abs/cond-mat/0611687}{{\ttfamily
  cond-mat/0611687}}].

\bibitem{PhysRevB.86.045139}
Z.Y.~Xie, J.~Chen, M.P.~Qin, J.W.~Zhu, L.P.~Yang and T.~Xiang,
  \emph{Coarse-graining renormalization by higher-order singular value
  decomposition}, \href{https://doi.org/10.1103/PhysRevB.86.045139}{\emph{Phys.
  Rev. B} {\bfseries 86} (2012) 045139}
  [\href{https://arxiv.org/abs/1201.1144}{{\ttfamily 1201.1144}}].

\bibitem{PhysRevLett.115.180405}
G.~Evenbly and G.~Vidal, \emph{Tensor network renormalization},
  \href{https://doi.org/10.1103/PhysRevLett.115.180405}{\emph{Phys. Rev. Lett.}
  {\bfseries 115} (2015) 180405}.

\bibitem{PhysRevB.97.045111}
M.~Hauru, C.~Delcamp and S.~Mizera, \emph{Renormalization of tensor networks
  using graph-independent local truncations},
  \href{https://doi.org/10.1103/PhysRevB.97.045111}{\emph{Phys. Rev. B}
  {\bfseries 97} (2018) 045111}.

\bibitem{PhysRevB.102.054432}
D.~Adachi, T.~Okubo and S.~Todo, \emph{Anisotropic tensor renormalization
  group}, \href{https://doi.org/10.1103/PhysRevB.102.054432}{\emph{Phys. Rev.
  B} {\bfseries 102} (2020) 054432}.

\bibitem{PhysRevB.105.L060402}
D.~Adachi, T.~Okubo and S.~Todo, \emph{Bond-weighted tensor renormalization
  group}, \href{https://doi.org/10.1103/PhysRevB.105.L060402}{\emph{Phys. Rev.
  B} {\bfseries 105} (2022) L060402}
  [\href{https://arxiv.org/abs/2011.01679}{{\ttfamily 2011.01679}}].

\bibitem{Nakayama:2023ytr}
K.~Nakayama, \emph{{Randomized higher-order tensor renormalization group}},
  \href{https://arxiv.org/abs/2307.14191}{{\ttfamily 2307.14191}}.

\bibitem{PhysRevLett.53.637}
N.~Seiberg, \emph{Topology in strong coupling},
  \href{https://doi.org/10.1103/PhysRevLett.53.637}{\emph{Phys. Rev. Lett.}
  {\bfseries 53} (1984) 637}.

\bibitem{PhysRevD.55.3966}
J.C.~Plefka and S.~Samuel, \emph{Strong-coupling analysis of the lattice
  {{${\mathrm{CP}}^{N\ensuremath{-}1}$}} models in the presence of a
  \ensuremath{\theta} term},
  \href{https://doi.org/10.1103/PhysRevD.55.3966}{\emph{Phys. Rev. D}
  {\bfseries 55} (1997) 3966}
  [\href{https://arxiv.org/abs/hep-lat/9612004}{{\ttfamily hep-lat/9612004}}].

\bibitem{HALDANE1983464}
F.~Haldane, \emph{Continuum dynamics of the 1-d heisenberg antiferromagnet:
  Identification with the {{O(3)}} nonlinear sigma model},
  \href{https://doi.org/https://doi.org/10.1016/0375-9601(83)90631-X}{\emph{Phys.
  Lett. A} {\bfseries 93} (1983) 464}.

\bibitem{PhysRevLett.50.1153}
F.D.M.~Haldane, \emph{Nonlinear field theory of large-spin heisenberg
  antiferromagnets: Semiclassically quantized solitons of the one-dimensional
  easy-axis n\'eel state},
  \href{https://doi.org/10.1103/PhysRevLett.50.1153}{\emph{Phys. Rev. Lett.}
  {\bfseries 50} (1983) 1153}.

\bibitem{PhysRevB.36.5291}
I.~Affleck and F.D.M.~Haldane, \emph{Critical theory of quantum spin chains},
  \href{https://doi.org/10.1103/PhysRevB.36.5291}{\emph{Phys. Rev. B}
  {\bfseries 36} (1987) 5291}.

\bibitem{PhysRevLett.66.2429}
I.~Affleck, \emph{Nonlinear \ensuremath{\sigma} model at
  \ensuremath{\theta}=\ensuremath{\pi}: Euclidean lattice formulation and
  solid-on-solid models},
  \href{https://doi.org/10.1103/PhysRevLett.66.2429}{\emph{Phys. Rev. Lett.}
  {\bfseries 66} (1991) 2429}.

\bibitem{Wess:1971yu}
J.~Wess and B.~Zumino, \emph{{Consequences of anomalous Ward identities}},
  \href{https://doi.org/10.1016/0370-2693(71)90582-X}{\emph{Phys. Lett. B}
  {\bfseries 37} (1971) 95}.

\bibitem{Witten:1983ar}
E.~Witten, \emph{{Nonabelian Bosonization in Two-Dimensions}},
  \href{https://doi.org/10.1007/BF01215276}{\emph{Commun. Math. Phys.}
  {\bfseries 92} (1984) 455}.

\bibitem{PhysRevLett.75.4524}
W.~Bietenholz, A.~Pochinsky and U.J.~Wiese, \emph{Meron-cluster simulation of
  the $\ensuremath{\theta}$ vacuum in the 2d {{O(3)}} model},
  \href{https://doi.org/10.1103/PhysRevLett.75.4524}{\emph{Phys. Rev. Lett.}
  {\bfseries 75} (1995) 4524}
  [\href{https://arxiv.org/abs/hep-lat/9505019}{{\ttfamily hep-lat/9505019}}].

\bibitem{PhysRevD.77.056008}
B.~All\'es and A.~Papa, \emph{Mass gap in the 2d {{O(3)}} nonlinear sigma model
  with a $\ensuremath{\theta}=\ensuremath{\pi}$ term},
  \href{https://doi.org/10.1103/PhysRevD.77.056008}{\emph{Phys. Rev. D}
  {\bfseries 77} (2008) 056008}
  [\href{https://arxiv.org/abs/0711.1496}{{\ttfamily 0711.1496}}].

\bibitem{PhysRevD.86.096009}
V.~Azcoiti, G.~Di~Carlo, E.~Follana and M.~Giordano, \emph{Critical behavior of
  the {{O(3)}} nonlinear sigma model with topological term at
  $\ensuremath{\theta}\mathbf{=}\ensuremath{\pi}$ from numerical simulations},
  \href{https://doi.org/10.1103/PhysRevD.86.096009}{\emph{Phys. Rev. D}
  {\bfseries 86} (2012) 096009}
  [\href{https://arxiv.org/abs/1207.4905}{{\ttfamily 1207.4905}}].

\bibitem{Caspar:2022llo}
S.~Caspar and H.~Singh, \emph{{From Asymptotic Freedom to {\ensuremath{\theta}}
  Vacua: Qubit Embeddings of the O(3) Nonlinear {\ensuremath{\sigma}} Model}},
  \href{https://doi.org/10.1103/PhysRevLett.129.022003}{\emph{Phys. Rev. Lett.}
  {\bfseries 129} (2022) 022003}
  [\href{https://arxiv.org/abs/2203.15766}{{\ttfamily 2203.15766}}].

\bibitem{Sulejmanpasic:2020lyq}
T.~Sulejmanpasic, D.~G{\"o}schl and C.~Gattringer, \emph{{First-Principles
  Simulations of 1+1D Quantum Field Theories at $\theta=\pi$ and Spin Chains}},
  \href{https://doi.org/10.1103/PhysRevLett.125.201602}{\emph{Phys. Rev. Lett.}
  {\bfseries 125} (2020) 201602}
  [\href{https://arxiv.org/abs/2007.06323}{{\ttfamily 2007.06323}}].

\bibitem{PhysRevLett.98.257203}
V.~Azcoiti, G.~Di~Carlo and A.~Galante, \emph{Critical behavior of
  {{${\mathrm{CP}}^{1}$}} at $\ensuremath{\theta}=\ensuremath{\pi}$,
  {{Haldane's}} conjecture, and the relevant universality class},
  \href{https://doi.org/10.1103/PhysRevLett.98.257203}{\emph{Phys. Rev. Lett.}
  {\bfseries 98} (2007) 257203}
  [\href{https://arxiv.org/abs/0710.1507}{{\ttfamily 0710.1507}}].

\bibitem{10.1143/ptp/93.1.161}
A.~Sayed~Hassan, M.~Imachi and H.~Yoneyama, \emph{{Real Space Renormalization
  Group Analysis of U(1)-Gauge Theory with $\theta$ Term in 2 Dimensions}},
  \href{https://doi.org/10.1143/ptp/93.1.161}{\emph{Progress of Theoretical
  Physics} {\bfseries 93} (1995) 161}
  [\href{https://arxiv.org/abs/https://academic.oup.com/ptp/article-pdf/93/1/161/5249019/93-1-161.pdf}{{\ttfamily
  https://academic.oup.com/ptp/article-pdf/93/1/161/5249019/93-1-161.pdf}}].

\bibitem{GENZ2003187}
A.~Genz, \emph{Fully symmetric interpolatory rules for multiple integrals over
  hyper-spherical surfaces},
  \href{https://doi.org/https://doi.org/10.1016/S0377-0427(03)00413-8}{\emph{J.
  Comput. App. Maths.} {\bfseries 157} (2003) 187}.

\bibitem{Kuramashi:2019cgs}
Y.~Kuramashi and Y.~Yoshimura, \emph{{Tensor renormalization group study of
  two-dimensional U(1) lattice gauge theory with a $\theta$ term}},
  \href{https://doi.org/10.1007/JHEP04(2020)089}{\emph{JHEP} {\bfseries 04}
  (2020) 089} [\href{https://arxiv.org/abs/1911.06480}{{\ttfamily
  1911.06480}}].

\bibitem{PhysRevB.48.16814}
K.~Nomura, \emph{Logarithmic corrections of the one-dimensional s=1/2
  heisenberg antiferromagnet},
  \href{https://doi.org/10.1103/PhysRevB.48.16814}{\emph{Phys. Rev. B}
  {\bfseries 48} (1993) 16814}.

\bibitem{PhysRevLett.55.1355}
I.~Affleck, \emph{Critical behavior of two-dimensional systems with continuous
  symmetries}, \href{https://doi.org/10.1103/PhysRevLett.55.1355}{\emph{Phys.
  Rev. Lett.} {\bfseries 55} (1985) 1355}.

\bibitem{KNomura_1994}
K.~Nomura and K.~Okamoto, \emph{Critical properties of s= 1/2 antiferromagnetic
  xxz chain with next-nearest-neighbour interactions},
  \href{https://doi.org/10.1088/0305-4470/27/17/012}{\emph{Journal of Physics
  A: Mathematical and General} {\bfseries 27} (1994) 5773}.

\bibitem{PhysRevB.104.165132}
A.~Ueda and M.~Oshikawa, \emph{Resolving the berezinskii-kosterlitz-thouless
  transition in the two-dimensional xy model with tensor-network-based level
  spectroscopy}, \href{https://doi.org/10.1103/PhysRevB.104.165132}{\emph{Phys.
  Rev. B} {\bfseries 104} (2021) 165132}.

\bibitem{PhysRevB.80.155131}
Z.-C.~Gu and X.-G.~Wen, \emph{Tensor-entanglement-filtering renormalization
  approach and symmetry-protected topological order},
  \href{https://doi.org/10.1103/PhysRevB.80.155131}{\emph{Phys. Rev. B}
  {\bfseries 80} (2009) 155131}
  [\href{https://arxiv.org/abs/0903.1069}{{\ttfamily 0903.1069}}].

\bibitem{DiFrancesco:1997nk}
P.~Di~Francesco, P.~Mathieu and D.~Senechal, \emph{{Conformal Field Theory}},
  Graduate Texts in Contemporary Physics, Springer-Verlag, New York (1997),
  \href{https://doi.org/10.1007/978-1-4612-2256-9}{10.1007/978-1-4612-2256-9}.

\bibitem{Ginsparg:1988ui}
P.H.~Ginsparg, \emph{{APPLIED CONFORMAL FIELD THEORY}},  in \emph{{Les Houches
  Summer School in Theoretical Physics: Fields, Strings, Critical Phenomena}},
  9, 1988 [\href{https://arxiv.org/abs/hep-th/9108028}{{\ttfamily
  hep-th/9108028}}].

\bibitem{CARDY1986186}
J.L.~Cardy, \emph{Operator content of two-dimensional conformally invariant
  theories},
  \href{https://doi.org/https://doi.org/10.1016/0550-3213(86)90552-3}{\emph{Nuclear
  Physics B} {\bfseries 270} (1986) 186}.

\bibitem{JLCardy_1986}
J.L.~Cardy, \emph{Logarithmic corrections to finite-size scaling in strips},
  \href{https://doi.org/10.1088/0305-4470/19/17/008}{\emph{Journal of Physics
  A: Mathematical and General} {\bfseries 19} (1986) L1093}.

\bibitem{JMKosterlitz_1973}
J.M.~Kosterlitz and D.J.~Thouless, \emph{Ordering, metastability and phase
  transitions in two-dimensional systems},
  \href{https://doi.org/10.1088/0022-3719/6/7/010}{\emph{Journal of Physics C:
  Solid State Physics} {\bfseries 6} (1973) 1181}.

\bibitem{JMKosterlitz_1974}
J.M.~Kosterlitz, \emph{The critical properties of the two-dimensional xy
  model}, \href{https://doi.org/10.1088/0022-3719/7/6/005}{\emph{Journal of
  Physics C: Solid State Physics} {\bfseries 7} (1974) 1046}.

\bibitem{Affleck:1984ar}
I.~Affleck, \emph{{The Quantum Hall Effect, $\sigma$ Models at $\theta = \pi$
  and Quantum Spin Chains}},
  \href{https://doi.org/10.1016/0550-3213(85)90353-0}{\emph{Nucl. Phys. B}
  {\bfseries 257} (1985) 397}.

\bibitem{Affleck:1985wb}
I.~Affleck, \emph{{Exact Critical Exponents for Quantum Spin Chains, Nonlinear
  Sigma Models at Theta = pi and the Quantum Hall Effect}},
  \href{https://doi.org/10.1016/0550-3213(86)90167-7}{\emph{Nucl. Phys. B}
  {\bfseries 265} (1986) 409}.

\bibitem{IAffleck_1989}
I.~Affleck, D.~Gepner, H.J.~Schulz and T.~Ziman, \emph{Critical behaviour of
  spin-s heisenberg antiferromagnetic chains: analytic and numerical results},
  \href{https://doi.org/10.1088/0305-4470/22/5/015}{\emph{Journal of Physics A:
  Mathematical and General} {\bfseries 22} (1989) 511}.

\bibitem{OKAMOTO1992433}
K.~Okamoto and K.~Nomura, \emph{Fluid-dimer critical point in s = 12
  antiferromagnetic heisenberg chain with next nearest neighbor interactions},
  \href{https://doi.org/https://doi.org/10.1016/0375-9601(92)90823-5}{\emph{Physics
  Letters A} {\bfseries 169} (1992) 433}.

\bibitem{Ueda:2023smj}
A.~Ueda and M.~Oshikawa, \emph{{Finite-size and finite bond dimension effects
  of tensor network renormalization}},
  \href{https://doi.org/10.1103/PhysRevB.108.024413}{\emph{Phys. Rev. B}
  {\bfseries 108} (2023) 024413}
  [\href{https://arxiv.org/abs/2302.06632}{{\ttfamily 2302.06632}}].

\bibitem{LUSCHER1991221}
M.~Luscher, P.~Weisz and U.~Wolff, \emph{A numerical method to compute the
  running coupling in asymptotically free theories},
  \href{https://doi.org/https://doi.org/10.1016/0550-3213(91)90298-C}{\emph{Nuclear
  Physics B} {\bfseries 359} (1991) 221}.

\bibitem{DAdda:1978vbw}
A.~D'Adda, M.~Luscher and P.~Di~Vecchia, \emph{{A 1/n Expandable Series of
  Nonlinear Sigma Models with Instantons}},
  \href{https://doi.org/10.1016/0550-3213(78)90432-7}{\emph{Nucl. Phys. B}
  {\bfseries 146} (1978) 63}.

\bibitem{Witten:1978bc}
E.~Witten, \emph{{Instantons, the Quark Model, and the 1/n Expansion}},
  \href{https://doi.org/10.1016/0550-3213(79)90243-8}{\emph{Nucl. Phys. B}
  {\bfseries 149} (1979) 285}.

\bibitem{Morita:2024lwg}
S.~Morita and N.~Kawashima, \emph{{Multi-impurity method for the bond-weighted
  tensor renormalization group}},
  \href{https://doi.org/10.1103/PhysRevB.111.054433}{\emph{Phys. Rev. B}
  {\bfseries 111} (2025) 054433}
  [\href{https://arxiv.org/abs/2411.13998}{{\ttfamily 2411.13998}}].

\end{thebibliography}\endgroup
\bibliographystyle{JHEP}
\end{document}